\documentclass[twocolumn,aps,epsfig,nofootinbib,preprintnumbers]{revtex4}
\usepackage{graphicx}
\usepackage{epstopdf}
\usepackage{latexsym}
\usepackage{amssymb}
\usepackage{amsmath}
\usepackage{color}
\usepackage{float}
\usepackage{mathrsfs}
\usepackage{multirow}
\usepackage{tabularx,booktabs}
\usepackage{esvect}
\usepackage{ulem}
\newcolumntype{Y}{>{\centering\arraybackslash}X}

\usepackage[center]{subfigure}
\usepackage{makecell}
\usepackage[colorlinks=true]{hyperref} 
\setcellgapes{4pt}

 \allowdisplaybreaks

\begin{document}

 \newcommand{\hsp}{\hspace{0.1mm}}
 \newcommand{\bq}{\begin{equation}}
 \newcommand{\eq}{\end{equation}}
 \newcommand{\bqn}{\begin{eqnarray}}
 \newcommand{\eqn}{\end{eqnarray}}
 \newcommand{\nb}{\nonumber}
 \newcommand{\lb}{\label}
 \newcommand{\pp}{\partial}

 \preprint{YITP-21-11, IPMU21-0011}

\title{Gravitational wave cosmology I:   high frequency  approximation}

\date{\today}

\author{Jared Fier$^{1}$}
\email{Jared$\_$Fier@baylor.edu}

\author{Xiongjun Fang$^{2}$}
\email{fangxj@hunnu.edu.cn}

\author{Bowen Li$^{1}$}
\email{Bowen$\_$Li@baylor.edu}
 
\author{Shinji Mukohyama$^{3, 4}$}
\email{shinji.mukohyama@yukawa.kyoto-u.ac.jp} 

\author{Anzhong Wang$^{1}$}
\email{Anzhong$\_$Wang@baylor.edu; corresponding author}

\author{Tao Zhu$^{5,6}$}
\email{zhut05@zjut.edu.cn}

\affiliation{
$^{1}$ GCAP-CASPER, Department of Physics, Baylor University, One Bear Place $\#$97316, Waco, Texas, 76798-7316, USA\\
$^{2}$ Department of Physics, Key Laboratory of Low Dimensional Quantum Structures and
Quantum Control of Ministry of Education, and Synergetic Innovation Center for Quantum Effects and Applications, Hunan Normal
University, Changsha, Hunan 410081, P. R. China\\
$^{3}$ Center for Gravitational Physics, Yukawa Institute for Theoretical Physics, Kyoto University, 606-8502, Kyoto, Japan \\
 $^{4}$ Kavli Institute for the Physics and Mathematics of the Universe (WPI), The University of Tokyo Institutes for Advanced Study,
 The University of Tokyo, Kashiwa, Chiba 277-8583, Japan\\
 ${}^{5}$Institute for Theoretical Physics \& Cosmology, Zhejiang University of Technology, Hangzhou, 310023, China\\
${}^{6}$ United Center for Gravitational Wave Physics (UCGWP),  Zhejiang University of Technology, Hangzhou, 310023, China
}

\begin{abstract}

In this paper, we systematically study gravitational waves (GWs) first produced by remote compact astrophysical sources  and then propagating in our 
inhomogeneous universe  through cosmic distances, before arriving at detectors. To describe such GWs properly, we introduce three scales, 
$\lambda, \; L_c$ and $L$, denoting, respectively, the typical wavelength of GWs,  the scale of the cosmological perturbations, and 
the size of the observable universe. For GWs to be detected by the current and foreseeable detectors, the condition $\lambda \ll L_c \ll L$ 
holds. Then, such GWs can be approximated as high-frequency GWs and be well separated from the background $\gamma_{\mu\nu}$ by 
averaging the spacetime curvatures over a scale $\ell$, where $\lambda \ll \ell \ll L_c $, and $g_{\mu\nu} = \gamma_{\mu\nu} + \epsilon 
h_{\mu\nu}$ with  $h_{\mu\nu}$ denoting the GWs. In order for the backreaction of the GWs to the background 
spacetimes to be negligible, we must assume that  $\left|h_{\mu\nu}\right| \ll 1$, in addition to the condition $\epsilon \ll 1$, which are 
also the conditions for the linearized Einstein field equations for $h_{\mu\nu}$ to be valid. Such studies can be significantly simplified by properly 
choosing gauges, such as the spatial, traceless, and Lorenz gauges. We show that these three different gauge conditions  can be imposed 
simultaneously, even when the background is not vacuum, as
long as the high-frequency GW approximation is valid. However, to develop the formulas that can be applicable to as many cases as possible, 
in this paper we first write down explicitly the linearized Einstein field equations by imposing only the spatial gauge.  Then, applying 
these formulas together with the geometrical optics approximation to such GWs,  we find that they still move along null geodesics and its polarization 
bi-vector is parallel-transported, even when both the cosmological scalar and tensor perturbations are present. In addition, we also calculate the 
gravitational integrated Sachs-Wolfe effects due to these two kinds of  perturbations, whereby the dependences of the amplitude, phase and luminosity 
distance of the GWs on these perturbations are read out explicitly.

\end{abstract}

\maketitle

\section{Introduction}
\renewcommand{\theequation}{1.\arabic{equation}} \setcounter{equation}{0}

The detection of the first gravitational wave (GW) from the coalescence of two massive black holes (BHs) by the advanced Laser Interferometer Gravitational-Wave Observatory 
(aLIGO) marked the beginning of a new era,  {\it the GW astronomy} \cite{Ref1}. Following this observation, soon more than 50 GWs were  detected by the LIGO/Virgo scientific 
collaboration \cite{GWTC1,GWTC2,aLIGO}. The outbreak of interest on GWs and BHs has further gained momenta after the detection of the  shadow of the M87 BH  
\cite{EHTa,EHTb,EHTc,EHTd,EHTe,EHTf}.

One of the remarkable observational results  is the discovery that the mass of  an individual BH in these binary systems can be  much larger than what was previously expected, 
both theoretically and observationally \cite{Ref4,Ref5,Ref6}, leading to the proposal and refinement of various formation scenarios, see, for example, \cite{Ref7,Ref8,SSTY18,LIGO-Virgo20}, and references therein. 
A consequence of this discovery 
is that the early inspiral phase may also be detectable by space-based observatories, such as LISA \cite{PAS17}, TianQin \cite{TianQin}, Taiji \cite{Taiji}, and DECIGO \cite{DECIGO}, 
for several years prior to their coalescence \cite{AS16, Moore15}. Multiple observations with different detectors at different frequencies of signals from the same source can provide excellent 
opportunities to study the evolution of the binary in detail. Since different detectors observe at disjoint frequency bands, together they cover different evolutionary stages of the same binary system. 
Each stage of the evolution carries information about different physical aspects of the source. As a result, multi-band GW detections will provide  an unprecedented opportunity to test  
different theories of gravity in the strong field regime \cite{Carson:2019kkh}.

Recently, some of the present authors generalized the post-Newtonian (PN) formalism to certain modified theories of gravity  and applied it to the quasi-circular inspiral of compact binaries. In particular, we calculated in detail the waveforms, GW polarizations, response functions and energy losses due to gravitational radiation in Brans-Dicke (BD) theory \cite{Ref23},  screened modified gravity (SMG) \cite{Tan18,Ref25,Ref25b},
and gravitational theories  with parity violations   \cite{QZZW19,ZZQW20,ZLWZWWZ20,QZZW20} to the leading PN order, with which we then considered projected constraints from the third-generation detectors. Such studies have been further generalized to triple systems \cite{Kai19,Zhao19} in Einstein-aether ($\ae$-) theory \cite{JM04,Jacobson08,OMW18}. When applying such formulas to the first relativistic triple system discovered in 2014 \cite{Ransom14}, we studied the radiation power, and found that quadrupole emission has almost the same amplitude as that in general relativity (GR), but the dipole emission can be as large as the quadrupole emission. This can provide a promising window to place severe constraints on $\ae$-theory with multi-band GW observations \cite{Ref17,Carson:2019yxq}.

More recently, we revisited the problem of a binary system of non-spinning bodies in a quasi-circular inspiral within the framework of $\ae$-theory \cite{Foster06,Foster07,Yagi13,Yagi14,HYY15,GHLP18}, and provided the explicit expressions for the time-domain and frequency-domain waveforms, GW polarizations, and response functions for both ground- and space-based detectors in the PN approximation \cite{Zhang20}. In particular, we found that, when going beyond  the leading order in the PN approximation, the non-Einsteinian polarization modes contain terms that depend on both the first and  second harmonics of the orbital phase. With this in mind, we  calculated analytically the corresponding parameterized post-Einsteinian parameters, generalizing the existing framework to allow for different propagation speeds among scalar, vector and tensor modes, without assuming the magnitude of its coupling parameters, and meanwhile allowing the binary system to have relative motions with respect to the aether field. Such results will particularly allow for the easy construction of Einstein-aether templates that could be used in Bayesian tests of GR in the future.

It is remarkable to note that the space-based  detectors mentioned above, together with the current and forthcoming  ground-based
 ones, such as KAGRA \cite{KAGRA},  Voyager \cite{Voyager},   the Einstein Telescope (ET) \cite{ET} and Cosmic Explorer (CE) \cite{CE},
 are able to detect GWs emitted from  such systems as far as the redshift is about $z \simeq 100$ \cite{HE19} \footnote{{It must be noted that, 
 according to structure formations, the first stars/galaxies should be formed
 at $z \simeq 20$ \cite{LL09}.  However, primordial BHs can be formed from the collapse of large overdensities in the radiation-dominated universe, which can explain the massive BHs
 observed so far from BBHs \cite{Franciolini21}. For recent reviews on this topic, see, for example, \cite{{GK21},SSTY18} and references therein.}}, which will result in 
a variety of profound scientific consequences. In particular, GWs propagating over such long cosmic distances will carry valuable information not only about their sources, but also 
about  the detail of the cosmological expansion and inhomogeneities of the universe, whereby a completely new window to explore the universe by using GWs is opened, as so far 
our understanding of the universe almost all comes from observations of electromagnetic waves only (possibly with the important exceptions of  cosmic rays and neutrinos) \cite{LL09}.

In this paper, we shall generalize our above studies to the cases in which  the GWs are first generated by remote astrophysical sources and then propagate in  the inhomogeneous universe 
through cosmic distances before  arriving at  detectors, either space- and/or ground-based ones. 
 It should be noted that recently such studies have already attracted lots of attention, see, for example,  \cite{Belgacem19} and references therein.
 In particular, using Isaacson's high frequency GW formulas \cite{Isaacson68a,Isaacson68b}, Laguna {\it et al} studied the gravitational analogue of the   electromagnetic integrated Sachs-Wolf (iSW) effects 
 in cosmology, and  found that the phase, frequency, and amplitude  of the GWs experience iSW effects, in addition to the magnifications on the amplitude from gravitational lensing   \cite{LLSY10}. 
 More recently,  Bertacca {\it et al}  connected the results of Laguna {\it et al}  obtained in real space frame to the observed frame, by using the cosmic rulers formulas \cite{SJ12},
 whereby  the corrections to  the luminosity distance due to velocity, volume, lensing and gravitational potential effects were calculated \cite{BRBM17}.

On the other hand,  Bonvin  {\it et al} \cite{BCST17} studied the effects of the  universe on   the gravitational waveform, and found that  the acceleration of the Universe and the peculiar acceleration 
of the binary with respect to the observer distort the gravitational chirp
  signals from the simplest GR prediction, not only a mere time independent rescaling of the chirp mass, but also the intrinsic parameter estimations for  binaries visible by LISA.
  In particular, the effect due to the peculiar acceleration can be much larger than the one due to the Universe acceleration.
  Moreover, peculiar accelerations can introduce a bias in the estimation of parameters such as the time of coalescence and the individual masses of the binary. An error in the estimation of the time of coalescence made by LISA will have an impact on the prediction of the time at which the signal will be visible by ground based interferometers, for signals spanning both frequency bands.
  
 Moreover, the correlations of such GWs  with lensing fields from the cosmic microwave background and galaxies were studied \cite{MWS20}, 
 whereby a new window to explore our universe by gravitational weak lensing was proposed.
 
Lately, GWs propagating in the curved universe  has been further generalized to scalar-tensor theories \cite{Garoffolo19}, including Horndeski \cite{DFL20a,DFL20b,EZ20} and SMG \cite{EZ20} theories. 

However, it should be noted that in all these studies, the cosmological tensor perturbations have been neglected {(Except \cite{DFL20a,DFL20b}, in which the background 
 is arbitrary.).}  As observing the primordial GWs (the tensor perturbations) is one of the main goals in 
the current and forthcoming cosmological observations   \cite{CMB-S4}, in this paper  we shall consider the cosmological background that consists of both the scalar and tensor perturbations, but restrict ourselves only to  Einstein's theory, and leave the generalizations to other theories of gravity to other occasions. What we are planning to do in  the current paper are the following:

\begin{itemize}

 \item  First, to describe the  GWs propagating through the inhomogeneous universe from cosmic distances to observers properly, we first introduce three scales, $\lambda, \; L_c$ and $L$, which denote, respectively, the typical wavelength of GWs,  the scale of the cosmological perturbations, and  the size of the observable universe. For GWs to be detected by the current and foreseeable detectors, we find that
 the condition 
 \bq
 \lb{eq1.1}
 \lambda \ll L_c \ll L,
 \eq
  always holds. Then, such GWs can be approximated as {\it high-frequency GWs} \footnote{  {It should be noted that Pulsar Timing Arrays can detect GWs with wavelengths ranging from
  an astronomical unit to a parsec \cite{NANOGrav20}. For such detections, the high-frequency approximations might not be valid any more \cite{DCL21}. We wish to come back to this subject soon.}}, and be well separated from the background $\gamma_{\mu\nu}$ by 
averaging the spacetime curvatures over a scale $\ell$, where $\lambda \ll \ell \ll L_c$, and 
 the total metric of the spacetime is given by
\bqn
\lb{eq1.2}
 g_{\mu\nu} &=&  \gamma_{\mu\nu} + \epsilon h_{\mu\nu}, 
 \eqn
 where   {$\epsilon \simeq {\cal{O}}(\lambda/L)$,} and $\gamma_{\mu\nu}$ denotes the background, while $h_{\mu\nu}$ represents the GWs.
 In order for the backreaction of the GWs to the background 
spacetimes to be negligible, we must assume that 
\bq
\lb{eq1.3}
\left|h_{\mu\nu}\right| \ll 1,
\eq
 in addition to the condition $\epsilon \ll 1$, which are  also the conditions for the linearized Einstein field equations for $h_{\mu\nu}$ to be valid.

\item  Such studies can be significantly simplified by properly 
imposing gauge conditions, such as {\it the spatial, traceless, and Lorenz gauges}, given, respectively, by
\bqn
\lb{eq1.1a}
\chi_{0\mu} &=& 0, \\
\lb{eq1.1b}
\chi&=& 0, \\
\lb{eq1.1c}
\nabla^{\nu}\chi_{\mu\nu} &=& 0, 
\eqn
where
\bq
\lb{eq1.4}
\chi_{\mu\nu} \equiv h_{\mu\nu} - \frac{1}{2}\gamma_{\mu\nu} h, \quad h \equiv  \gamma^{\mu\nu}h_{\mu\nu},
\eq
  and $\nabla^{\nu}$ denotes the covariant derivative with respect to 
$\gamma_{\mu\nu}$. We show that these three different gauge conditions  can be imposed simultaneously, even when the background is not vacuum, as
longer as the high-frequency GW approximations are valid.

\item  However, to develop the formulas that can be applicable to as many cases as possible, 
in this paper we write down explicitly the linearized Einstein field equations for $\chi_{\mu\nu}$ by imposing only the spatial gauge.  
 Applying  these formulas together with the geometrical optic approximations to such GWs,  {we find the well-known results \cite{MTW73}:}
   they still move along null geodesics and its polarization bi-vector is 
parallel-transported, even when both the cosmological scalar and tensor perturbations are present. In addition, we also calculate the gravitational 
integrated Sachs-Wolfe (iSW) effects due to these two kinds of  perturbations, whereby the dependences of the amplitude, phase and luminosity distance 
of the GWs on these perturbations are read off explicitly. 

\end{itemize}

 The rest of the paper is organized as follows:  In Sec. II,  after   introducing the three different scales, $\lambda, L_c, L$, we show that, for the GWs to be detected by the current 
 and foreseeable both ground- and space-based detectors,  such GWs can be well approximated as high frequency GWs. Then, we derive the Einstein field equations, and find that,
to make the backreaction of the GWs to the background negligible, as well as to have the linearized
Einstein field equations for  $h_{\mu\nu}$ to be valid, the condition  (\ref{eq1.3}) must hold. In this section, we also provide a very brief review on the cosmological background
that consists of both the cosmological and tensor perturbations. 
 In Sec. III, we consider   the gauge freedom for  GWs, and show that the three different gauge conditions,  (\ref{eq1.1a})-(\ref{eq1.1c}), 
 can be still imposed simultaneously, even when the background spacetime is   not vacuum, as long as the high-frequency approximations are valid. 
 Then, by imposing only the spatial gauge condition (\ref{eq1.1a}), we write down the linearized Einstein field equations for the GWs, so the formulas can be applied to cases with different choices of gauges. 
In Sec. IV we study the GWs with the geometrical optics approximation, and calculate the effects of the cosmological scalar and tensor perturbations on the
 amplitudes and phases of such  GWs, and find the  explicit expressions   of the iSW effects due to both the cosmological scalar and tensor perturbations. 
 When applying them to a binary system, we calculate explicitly the effects of these two kinds of the cosmological perturbations on the  luminosity distance and the chirp mass [cf. Eq.(\ref{eq4.41})]. 
 Finally, we summarize our main results in Sec. V, and present some concluding remarks. 
 
 There are also three appendices, A, B and C, in which some mathematical computations are presented. In particular, in Appendix A, we give a very brief review over the inhomogeneous universe, 
 when both the cosmological scalar and tensor perturbations are present, while in Appendix B, we present the field equations for the GWs $\chi_{\mu\nu}$ by imposing only the spatial gauge (\ref{eq1.1a}).
 In Appendix C, we first decompose $\chi_{\mu\nu}$ as $\chi_{\mu\nu} = \chi_{\mu\nu}^{(0)} + \epsilon_c \chi_{\mu\nu}^{(1)}$ and then write down explicitly the field equations for $\chi_{\mu\nu}^{(1)}$ only with 
 the  spatial gauge.

Before proceeding to the next section, we would like to note that GWs produced by remote astrophysical sources and then propagating through the homogeneous and isotropic universe have been 
systematical studied by Ashtekar and his collaborators through a series of papers \cite{AshtekarI,AshtekarII,AshtekarIII,AshtekarIIIa,AshtekarIV,Ashtekar20a,Ashtekar20b}, and various subtle issues related to the de Sitter background were clarified \cite{Ashtekar17a,Ashtekar17b,Ashtekar17c} (See also \cite{Chu15,Bishop16,TW16,DH16,BGY17a,BGY17b,Chu17,HFZ19,FSW20}).

In addition, in this paper we shall adopt the following conventions, which are  different from those adopted in
\cite{Isaacson68a,Isaacson68b}, but the same as those used in \cite{MM16}. In particular, in this paper the signature of the metric is 
($-, +, +, +$), while the Christoffel symbols, Riemann and Ricci tensors, as well as the Ricci scalar,  are defined, respectively, by
\bqn
\lb{eq1.a}
&& \Gamma^\alpha_{\mu\nu} \equiv \frac{1}{2}g^{\alpha\beta}\left(g_{\beta\nu, \mu} + g_{\beta\mu, \nu} - g_{\mu\nu, \beta}\right),\nb\\
&& \left(D_{\alpha}D_{\beta} - D_{\beta}D_{\alpha}\right) X^{\mu} = {R^{\mu}}_{\nu\alpha\beta}X^{\nu}, \nb\\
&& R_{\mu\nu} \equiv R^{\alpha}_{\;\;\mu\alpha\nu}, \quad R \equiv g^{\mu\nu} R_{\mu\nu},
\eqn
where $D_{\alpha}$ denotes the covariant derivative with respect to  metric $g_{\mu\nu}$, $g_{\mu\nu,\lambda} \equiv \partial g_{\mu\nu}/\partial x^{\lambda}$, and
\bqn
\lb{eq1.b}
R^{\alpha}_{\;\;\mu\nu\lambda} &=& \Gamma^{\alpha}_{\mu\lambda,\nu}- \Gamma^{\alpha}_{\mu\nu,\lambda}
+ \Gamma^{\alpha}_{\beta\nu}\Gamma^{\beta}_{\mu\lambda} - \Gamma^{\alpha}_{\beta\lambda}\Gamma^{\beta}_{\mu\nu}.
\eqn
The Einstein field equations read,
\bqn
\lb{eq1.c}
R_{\mu\nu} - \frac{1}{2}g_{\mu\nu} R = \kappa T_{\mu\nu},
\eqn
where $\kappa \equiv 8\pi G/c^4$, with $G$ denoting the Newtonian constant, and $c$ the speed of light. In addition to $D_{\alpha}$ and $\nabla_{\alpha}$, we also
introduce the covariant derivative $\bar\nabla_{\alpha}$ with respect to the homogeneous metric $\bar{\gamma}_{\mu\nu}$, where
\bqn
\lb{eq1.5}
 \gamma_{\mu\nu} &=& \bar\gamma_{\mu\nu} + \epsilon_c \hat\gamma_{\mu\nu},
 \eqn
with {$\epsilon_c \simeq {\cal{O}}(L_c/L) \ll 1$.} We shall also adopt  the conventions,
$A_{(\mu\nu)} \equiv \left(A_{\mu\nu} + A_{\nu\mu} \right)/2, \; A_{[\mu\nu]} \equiv \left(A_{\mu\nu} - A_{\nu\mu} \right)/2$.

\section{Gravitational waves   Propagating  in    Inhomogeneous Universe}
\renewcommand{\theequation}{2.\arabic{equation}} \setcounter{equation}{0}

In this section, we shall consider   GWs first produced by  remote astrophysical sources  and then propagating in cosmic distances through the 
 inhomogeneous  Universe, before arriving at detectors.  To study such GWs, let us first consider several characteristic lengths that are highly relevant to
 their  propagations and polarizations.

\subsection{Characteristic Scales of Background}

In this paper, we shall consider our inhomogeneous  universe as the background, which includes two parts, the homogeneous and isotropic universe and its
inhomogeneous perturbations, given by $\bar{\gamma}_{\mu\nu}$ and $\hat{\gamma}_{\mu\nu}$, respectively, so the background metric ${\gamma}_{\mu\nu}$
can be written as
\bqn
\lb{eq2.1aa}
 \gamma_{\mu\nu} &=& \bar\gamma_{\mu\nu} + \epsilon_c \hat\gamma_{\mu\nu} + {\cal{O}}\left(\epsilon_c^2\right),\nb\\
 \gamma^{\mu\nu} &=& \bar\gamma^{\mu\nu} - \epsilon_c \hat\gamma^{\mu\nu} + {\cal{O}}\left(\epsilon_c^2\right),
 \eqn
where    $\epsilon_c, \; \left|\hat\gamma\right|   \ll 1$ [cf. Eq.(\ref{eq2.29h})],  and 
\bqn
\lb{eq2.2aa}
&& \gamma^{\mu\lambda}  \gamma_{\nu\lambda} = \delta^{\mu}_{\nu} +    {\cal{O}}\left(\epsilon_c^2\right),\quad
\bar\gamma^{\mu\lambda}  \bar\gamma_{\nu\lambda} = \delta^{\mu}_{\nu} +    {\cal{O}}\left(\epsilon_c^2\right),\nb\\
&&  \hat\gamma^{\mu}_{\nu} \equiv \bar\gamma^{\mu\alpha} \hat\gamma_{\alpha\nu}, \quad
\hat\gamma^{\mu\nu} \equiv \bar\gamma^{\mu\alpha} \bar\gamma^{\nu\beta} \hat\gamma_{\alpha\beta}, 
\eqn
and so on.

 The size of the observational universe is about  $L \simeq 8.8 \times 10^{26}\; m$ \cite{Gott05}. On the other hand, in the momentum space of the cosmological perturbations, 
 we have $L_c \simeq 1/k$, where $k$ denotes the typical wavenumber
of the perturbations, and $L_c$ the length over which the change of the cosmological perturbations becomes appreciable. When the
modes are outside the Hubble horizon, it can be shown that  $L_c/L \simeq 10^{-5}$. But, once they re-enter  the horizon these modes decay suddenly and then
are oscillating rapidly about a minimum \cite{SD03}.  In addition,  {the current temperature anisotropy $\Delta T/ T_0$ of the universe  is of order   $10^{-5}$
\cite{BMM04}. }
So, it is quite reasonable to assume that 
\bq
\lb{eq2.4aa}
\epsilon_c \simeq \frac{L_c}{L}   \ll 1. 
\eq

\subsection{Typical Gravitational Wavelengths}

For the second generation of the ground-based detectors, such as LIGO, Virgo, and KAGRA, the wavelength of the
detected GWs are $ \lambda   \simeq  10^{5}  \sim  10^{7}\;  m$, while the wavelength of
GWs to be detected by the space-based detectors, such as LISA, TianQin and Taiji, are  $\lambda  \simeq  10^{8}  \sim  10^{12}\;  m$
\footnote{The frequencies of GWs detected by the second generation of the ground-based detectors
is $f \simeq 20 - 2000$ Hz, while  the frequencies of GWs to be detected by  the space-based detectors are $f \simeq 1 - 10^{-4}$ Hz.}.
  Therefore, for the ground-based
detectors, we have $\epsilon \simeq \lambda/L \in \left(10^{-22}, \; 10^{-20}\right)$, while for the space-based detectors, we have  $\epsilon  \in \left(10^{-19}, \; 10^{-15}\right)$.

Therefore,  {in this paper we shall consider only the cases in which  the following is  true,
\bq
\lb{eq2.13b}
\frac{\lambda}{L_c} = \frac{\epsilon}{\epsilon_c} \ll 1,
\eq
so that  all GWs considered in this paper can be well approximated  as high frequency GWs.}

\subsection{Einstein Field Equations}

Following the above analyses,  we find that $\lambda$,  $L_c$ and $L$ denote, respectively,  the characteristic length over which $h_{\mu\nu}$,  $\hat\gamma_{\mu\nu}$ 
or $\bar\gamma_{\mu\nu}$ changes significantly. Thus, their derivatives are typically of the orders,
\bqn
\lb{eq2.1a}
&& \partial \bar\gamma  \sim \frac{\bar\gamma}{L}, \quad \partial^2 \bar\gamma  \sim \frac{\bar\gamma}{L^2}, \nb\\
&& \partial \hat\gamma  \sim \frac{\hat\gamma}{L_c}, \quad \partial^2 \hat\gamma  \sim \frac{\hat\gamma}{L_c^2},\nb\\
&& \partial h  \sim \frac{h}{\lambda}, \quad \partial^2 h  \sim  \frac{h}{\lambda^2}.
\eqn
To estimate orders of terms, following Isaacson \cite{Isaacson68a}, we  regard $L$ as order of unity, and
say that the metric (\ref{eq1.2}) contains a high-frequency GW, if and only if there exists
a family of coordinate systems (related by infinitesimal coordinate transformations), in which we have 
\bq
\lb{eq2.1}
\epsilon \ll \epsilon_c \ll 1, 
\eq
 and
\bqn
\lb{eq2.2}
&& \bar\gamma_{\mu\nu},  \; \bar\gamma_{\mu\nu,\alpha}, \; \bar\gamma_{\mu\nu,\alpha\beta} \simeq {\cal{O}}(1), \nb\\
 && \hat\gamma_{\mu\nu} \simeq {\cal{O}}\left(\hat\gamma\right), \;\; \hat\gamma_{\mu\nu,\alpha} \simeq {\cal{O}}\left(\hat\gamma/\epsilon_c\right), \nb\\
  &&   \hat\gamma_{\mu\nu,\alpha\beta} \simeq {\cal{O}}\left(\hat\gamma/\epsilon_c^2\right), \nb\\
 &&   h_{\mu\nu} \simeq {\cal{O}}\left(h\right), \;\;\; h_{\mu\nu,\alpha} \simeq {\cal{O}}\left(h/\epsilon\right),  \nb\\
  && h_{\mu\nu,\alpha\beta} \simeq {\cal{O}}\left(h/\epsilon^{2}\right),
 \eqn
 where $\gamma_{\mu\nu,\alpha} \equiv \partial \gamma_{\mu\nu}/\partial x^{\alpha}$, etc.
 Note that, in contrast to \cite{Isaacson68a}, here we do not assume $h_{\mu\nu} \simeq {\cal{O}}\left(1\right)$,
 in order to neglect the backreaction of the GWs to the background spacetime $\gamma_{\mu\nu}$, as to be shown below.

Expanding the Riemann and Ricci tensors $R_{\mu\nu\alpha\beta}\left(g_{\mu\nu}\right)$ and $R_{\mu\nu}\left(g_{\mu\nu}\right)$ in terms of $\epsilon$,
we find \cite{Isaacson68a,MM16},
\bqn
\lb{eq2.3}
 R_{\alpha\beta\gamma\delta}\left(g_{\mu\nu}\right) &=& {R_{\alpha\beta\gamma\delta}}^{(0)} + \epsilon {R_{\alpha\beta\gamma\delta}}^{(1)} + \epsilon^2 {R_{\alpha\beta\gamma\delta}}^{(2)}  \nb\\
&&   +  {\cal{O}}\left(\epsilon^{3}\right),\nb\\
R_{\alpha\beta}\left(g_{\mu\nu}\right) &=& {R_{\alpha\beta}}^{(0)} + \epsilon {R_{\alpha\beta}}^{(1)} + \epsilon^2 {R_{\alpha\beta}}^{(2)} \nb\\
&&     +  {\cal{O}}\left(\epsilon^{3}\right),
\eqn
where
\bqn
\lb{eq2.4}
&&  {R_{\alpha\beta\gamma\delta}}^{(0)} =  {R_{\alpha\beta\gamma\delta}}\left(\gamma_{\mu\nu}\right),\nb\\
&&   {R_{\alpha\beta\gamma\delta}}^{(1)} = \frac{1}{2}\Big[h_{\beta\gamma;\alpha\delta} + h_{\alpha\delta; \beta\gamma} - h_{\alpha\gamma; \beta\delta} - h_{\beta\delta; \alpha\gamma}
\nb\\
&& ~~~~~~~~~~~~~~~~~~~   {+    {R_{\alpha\sigma\gamma\delta}}^{(0)} h^{\sigma}_{\beta} - {R_{\beta\sigma\gamma\delta}}^{(0)} h^{\sigma}_{\alpha}}\Big],\\
\lb{eq2.4a}
&&  {R_{\alpha\beta}}^{(0)} =  {R_{\alpha\beta}}\left(\gamma_{\mu\nu}\right),\nb\\
&&   {R_{\alpha\beta}}^{(1)} = \frac{1}{2}\gamma^{\rho\tau}\Big(h_{\tau\alpha; \beta\rho} +  h_{\tau\beta; \alpha\rho}  \nb\\
&& ~~~~~~~~~~~~~~~~~~~~~~   - h_{\rho\tau;\alpha \beta} - h_{\alpha \beta; \rho\tau}\Big),\\
\lb{eq2.4ab}
&&   {R_{\alpha\beta}}^{(2)} = \frac{1}{4}\Big\{{h^{\rho\tau}}_{;\beta}h_{\rho\tau;\alpha} + 2h^{\rho\tau}\big(h_{\tau\rho;\alpha\beta} + h_{\alpha\beta; \tau\rho}\nb\\
&& ~~~~~~~~~~~~~~~~~ - h_{\tau\alpha;\beta\rho} - h_{\tau\beta;\alpha\rho}\big) + 2 {h^{\tau}_{\beta}}^{;\rho}\big(h_{\tau\alpha;\rho} - h_{\rho\alpha;\tau}\big)\nb\\
&& ~~~~~~~~~~~~~~~~~ - \left(2{h^{\rho\tau}}_{;\rho} - h^{;\tau}\right)\big(h_{\tau\alpha;\beta} + h_{\tau\beta;\alpha} - h_{\alpha\beta;\tau}\big)\Big\}. \nb\\
\eqn
Here the semi-colon ``;'' denotes the covariant derivative with respect to the background metric $\gamma_{\mu\nu}$.
For the sake of convenience, we shall also use $\nabla_{\lambda}$ to denote  the covariant derivative with respect to  $\gamma_{\mu\nu}$, so we have $h_{\mu\nu;\lambda} \equiv \nabla_{\lambda}h_{\mu\nu}$,
etc.  The background metric $\gamma_{\mu\nu}$ ($\gamma^{\mu\nu}$) is also used to lower (raise) the indices of $h_{\mu\nu}$, such as 
\bq
\lb{eq2.4b}
h^{\mu}_{\nu} \equiv \gamma^{\mu\alpha}h_{\alpha\nu} = \gamma_{\nu\alpha}h^{\mu\alpha}, \quad h   \equiv h^{\lambda}_{\lambda} =  \gamma^{\alpha\beta}h_{\alpha\beta},
\eq
and so on.

The background curvatures  ${R_{\alpha\beta\gamma\delta}}^{(0)}(\gamma)$ and ${R_{\alpha\beta}}^{(0)}(\gamma)$ can  be further expanded   in terms of $\epsilon_{c}$,  as
\bqn
\lb{eq2.15}
&& {R_{\alpha\beta\gamma\delta}}^{(0)}(\gamma) = \bar{R}_{\alpha\beta\gamma\delta}(\bar\gamma) + \epsilon_c \hat R_{\alpha\beta\gamma\delta}(\hat\gamma)\nb\\
&& ~~~~~~~~~~~~~~~~~~~  + \epsilon_c^2 \hat R^{(2)}_{\alpha\beta\gamma\delta}(\hat\gamma) +  {\cal{O}}\left(\epsilon_c^{3}\right),\nb\\
&& {R_{\alpha\beta}}^{(0)}(\gamma) = \bar R_{\alpha\beta}(\bar\gamma) + \epsilon_c \hat R_{\alpha\beta}(\hat\gamma) \nb\\
&& ~~~~~~~~~~~~~~~~~~~  + \epsilon_c^2 \hat R^{(2)}_{\alpha\beta}(\hat\gamma) +  {\cal{O}}\left(\epsilon_c^{3}\right), 
\eqn
where 
\bqn
\lb{eq2.16}
&&   \hat R_{\alpha\beta\gamma\delta} (\hat\gamma) = \frac{1}{2}\Big[\hat \gamma_{\beta\gamma|\alpha\delta} + \hat \gamma_{\alpha\delta|\beta\gamma}
- \hat \gamma_{\alpha\gamma| \beta\delta} - \hat \gamma_{\beta\delta| \alpha\gamma} \nb\\
&& ~~~~~~~~~~~~~~~~~~~~   {+ \bar R_{\alpha\sigma\gamma\delta} \hat \gamma^{\sigma}_{\beta} - \bar R_{\beta\sigma\gamma\delta} \hat \gamma^{\sigma}_{\alpha}}\Big],\\
&&   \hat R_{\alpha\beta} (\hat\gamma) = \frac{1}{2}\bar\gamma^{\rho\tau}\Big(\hat \gamma_{\tau\alpha| \beta\rho} + \hat \gamma_{\tau\beta| \alpha\rho}
- \hat \gamma_{\rho\tau|\alpha \beta} - \hat \gamma_{\alpha \beta| \rho\tau}\Big),\nb\\
\eqn
and   $\hat R^{(2)}_{\alpha\beta}(\hat\gamma)$ is given by Eq.(\ref{eq2.4ab})  with the replacement
$\left(h_{\alpha\beta}, \nabla_{\mu}\right) \rightarrow \left(\hat\gamma_{\alpha\beta}, \bar\nabla_{\mu}\right)$. 
Here  the vertical bar ``$|$'' denotes the covariant derivative with respect to $\bar\gamma_{\mu\nu}$, which is also denoted by $\bar\nabla_{\lambda}$,  so that  $\hat \gamma_{\rho\tau|\alpha}
\equiv \bar\nabla_{\alpha} \hat \gamma_{\rho\tau}$, etc. 
Taking $L \simeq  {\cal{O}}(1)$ and considering Eq.(\ref{eq2.2}) we find
\bqn
\lb{eq2.13ca}
&&  \bar R^{\alpha}_{\;\;\beta\gamma\delta}, \;  \bar R_{\alpha\beta}  \simeq {\cal{O}}(1),\\
\lb{eq2.13cb}
&& \epsilon_c \hat R^{\alpha}_{\;\; \beta\gamma\delta}, \;  \epsilon_c \hat R_{\alpha\beta}  \sim {\cal{O}}\left(\hat\gamma/\epsilon_c\right),\nb\\
   &&   \epsilon_c^2 {\hat R_{\alpha\beta\gamma\delta}}^{(2)},  \;   \epsilon^2_c {\hat R_{\alpha\beta}}^{(2)}  \simeq   {\cal{O}}\left(\hat\gamma^2\right),\\
\lb{eq2.13cc}
&&  \epsilon {R_{\alpha\beta\gamma\delta}}^{(1)},  \;   \epsilon {R_{\alpha\beta}}^{(1)}  \simeq   {\cal{O}}\left(h/\epsilon\right),\nb\\
   &&   \epsilon^2 {R_{\alpha\beta\gamma\delta}}^{(2)},  \;   \epsilon^2 {R_{\alpha\beta}}^{(2)}  \simeq   {\cal{O}}\left(h^2\right).
 \eqn

To write down the Einstein field equations, let us first note that
\bqn
\lb{eq2.5b}
 \left(\nabla_{\alpha}\nabla_{\beta}-\nabla_{\beta}\nabla_{\alpha}\right)\chi_{\gamma\delta}&=&-{R^{\sigma}_{\;\;\gamma\alpha\beta}}^{(0)} \chi_{\sigma\delta}\nb\\
 && -{R^{\sigma}_{\;\;\delta\alpha\beta}}^{(0)}\chi_{\gamma\sigma}.
 \eqn
 Then, we find that in terms of $\chi_{\mu\nu}$,  ${R_{\alpha\beta}}^{(1)}$ is given by
 \bqn
 \lb{eq2.5c}
  {R_{\alpha\beta}}^{(1)}&=&\frac{1}{2}\Big(2{R_{\gamma\alpha\beta\sigma}}^{(0)}\chi^{\gamma\sigma}+{R^{\sigma}_{\alpha}}^{(0)} \chi_{\beta\sigma} +{R^{\sigma}_{\beta}}^{(0)}\chi_{\alpha\sigma}\nb\\
  &&  ~~~~
  +\nabla_{\alpha}\nabla^{\delta}\chi_{\beta\delta}+\nabla_{\beta}\nabla^{\delta}\chi_{\alpha\delta}\Big) \nonumber \\
    &&-\frac{1}{2}\square\chi_{\alpha\beta}+\frac{1}{4}\gamma_{\alpha\beta}\square\chi,
 \eqn
  where  $\Box  \chi_{\alpha\beta}  \equiv \gamma^{\mu\nu}\chi_{\alpha\beta;\mu\nu}$, and
\bqn
\lb{eq2.11d}
\chi_{\mu\nu} \equiv h_{\mu\nu} - \frac{1}{2} \gamma_{\mu\nu} h, \quad \chi \equiv \gamma^{\mu\nu} \chi_{\mu\nu} = - h.
\eqn

It should be noted that in \cite{Isaacson68a} Isaacson considered the vacuum case, for which we have $ {R_{\alpha\beta}}^{(1)} = 0$,
 that is,
 \bqn
 \lb{eq2.5ca}
  &&\square\chi_{\alpha\beta} -\frac{1}{2}\gamma_{\alpha\beta}\square\chi -\nabla_{\alpha}\nabla^{\delta}\chi_{\beta\delta}-\nabla_{\beta}\nabla^{\delta}\chi_{\alpha\delta}
 \nb\\
  && ~~~~~~~  + 2{R_{\alpha\gamma\beta\sigma}}^{(0)}\chi^{\gamma\sigma}  - {R^{\sigma}_{\alpha}}^{(0)} \chi_{\beta\sigma} -{R^{\sigma}_{\beta}}^{(0)}\chi_{\alpha\sigma} =0,\nb\\
  && ~~~~~~~ ~~~~~~~ ~~~~~~~ ~~~~~~~ ~~~~~~~  
 \eqn
which is precisely Eq.(5.7) of \cite{Isaacson68a}, after the difference between the conventions used here and the ones used in \cite{Isaacson68a} is taken into account.

However, in the present paper we consider the propagation of GWs through the inhomogeneous universe, which has non-zero Riemann and Ricci tensors. 
So, we expect that  the corresponding Einstein field equations for $h_{\mu\nu}$ are different from Eq.(\ref{eq2.5ca}). To see this, we first note that
\bqn
\lb{eq2.5caa}
g^{\mu\nu} &=& \gamma^{\mu\nu} - \epsilon h^{\mu\nu} + \epsilon^2 h^{\mu}_{\alpha}h^{\alpha\nu} + O\left(\epsilon^3\right),\nb\\
R&\equiv& g^{\mu\nu}R_{\mu\nu} = R^{(0)} + \epsilon R^{(1)} + \epsilon^2 R^{(2)} + O\left(\epsilon^3\right), ~~~~~~~~
\eqn
where
\bqn
\lb{eq2.5cab}
R^{(0)} &\equiv&  \gamma^{\mu\nu}R^{(0)}_{\mu\nu}, \nb\\
R^{(1)} &\equiv&   \gamma^{\mu\nu}R^{(1)}_{\mu\nu} -  h^{\mu\nu}R^{(0)}_{\mu\nu}\nb\\
 &=& \nabla^{\alpha}\nabla^{\beta}\chi_{\alpha\beta}  -\chi^{\alpha\beta}R^{(0)}_{\alpha\beta} + \frac{1}{2}\left(\square + R^{(0)}\right) \chi, \nb\\
R^{(2)} &\equiv&   \gamma^{\mu\nu}R^{(2)}_{\mu\nu} -  h^{\mu\nu}R^{(1)}_{\mu\nu} +  h^{\mu}_{\alpha}h^{\alpha\nu}R^{(0)}_{\mu\nu}.
\eqn
Inserting Eqs.(\ref{eq2.3}) and (\ref{eq2.5caa}) into the Einstein field equations, we find that
\bqn
\lb{eq2.5cac}
R^{(0)}_{\mu\nu} &-&  \frac{1}{2}\gamma_{\mu\nu}R^{(0)} \nb\\
&+&  \epsilon\left[R^{(1)}_{\mu\nu} - \frac{1}{2}\left(\gamma_{\mu\nu}R^{(1)} + h_{\mu\nu}R^{(0)}\right)\right]\nb\\
&+&  \epsilon^2\left[R^{(2)}_{\mu\nu} - \frac{1}{2}\left(\gamma_{\mu\nu}R^{(2)} + h_{\mu\nu}R^{(1)}\right)\right] +  O\left(\epsilon^3\right)\nb\\
&=&  \kappa\left(T^{(0)}_{\mu\nu} +\epsilon{\cal{T}}_{\mu\nu}\right),  
\eqn
where $T^{(0)}_{\mu\nu}$ denote the energy-momentum tensor that produces the background, 
while ${\cal{T}}_{\mu\nu}$ denotes the astrophysical source that produces the GWs.

\subsection{Separation of GWs from Background}

   To separate   GWs produced by astrophysical sources  from the inhomogeneous background, we can average the field equations  over a length scale $\ell$,
 which is much larger than the typical wavelength of the GWs   but
much smaller than $L_c$,  
\bq
\lb{eq2.5g}
\lambda \ll \ell \ll L_c.
\eq
Then, this process will extract the slowly varying background from GWs, as the latter will vanish when averaging over such a scale. In particular, we have
\bqn
\lb{eq2.5ga}
&& \left<\gamma_{\mu\nu}\right> =\gamma_{\mu\nu}, \;\; \left<{R_{\mu\nu\alpha\beta}}^{(0)}\right> = {R_{\mu\nu\alpha\beta}}^{(0)}, \nb\\
&&  \left<{R_{\mu\nu}}^{(0)}\right> = {R_{\mu\nu}}^{(0)},\;\;  \left<{T_{\mu\nu}}^{(0)}\right> = {T_{\mu\nu}}^{(0)},\\
\lb{eq2.5gb}
&& \left<h_{\mu\nu}\right> =   \left<{R_{\mu\nu}}^{(1)}\right> =  \left<R^{(1)}\right>  = 0,\\
&&   \left<{R_{\mu\nu}}^{(2)}\right> =\left<{R_{\mu\nu}}^{(2)}\right>_{\ell}, \;\;  \left<R^{(2)}\right> =\left<R^{(2)}\right>_{\ell},\nb\\
&& \left<h_{\mu\nu}R^{(1)}\right> =\left<h_{\mu\nu}R^{(1)}\right>_{\ell},    \;\;
  \left<{{\cal{T}}_{\mu\nu}}\right> =\left<{{\cal{T}}_{\mu\nu}}\right>_{\ell}. ~~~~  
\eqn
Note that quadratic terms of $h_{\mu\nu}$ may survive such an averaging process, if two modes are almost equal but with different signs, although   each of them represents a high frequency mode.
For example, for $h_{\mu\nu} \propto e^{i\omega_1 x}$ and $h_{\alpha\beta} \propto e^{-i\omega_2 x}$, we have $h_{\mu\nu}h_{\alpha\beta} \propto e^{i\omega_{12} x}$, where
$\omega_{12} \equiv  \omega_1 - \omega_2$. Thus, although $\omega_1, \; \omega_2 \gg 1$, we can have $\omega_{12} \ll 1$, if  $\omega_1 \simeq \omega_{2}$. Therefore, due to the nonlinear
interactions among different modes, low frequency modes can be produced, which will survive with such averaging processes. If we are only interested in the linearized Einstein field equations 
of $h_{\mu\nu}$, such modes must be taken of care   properly.  With this in mind, taking the average of Eq.(\ref{eq2.5cac}) we find that 
\bqn
\lb{eq2.5gd}
R^{(0)}_{\mu\nu} -  \frac{1}{2}\gamma_{\mu\nu}R^{(0)} +  \epsilon^2  \left<{G_{\mu\nu}}^{(2)}\right>_{\ell} =   \kappa \left(T^{(0)}_{\mu\nu}+\epsilon \left<{{\cal{T}}_{\mu\nu}}\right>_{\ell}\right),\nb\\
\eqn
where 
\bqn
\lb{eq2.5ge}
 {G_{\mu\nu}}^{(2)} \equiv  R^{(2)}_{\mu\nu} - \frac{1}{2}\left(\gamma_{\mu\nu}R^{(2)} + h_{\mu\nu}R^{(1)}\right),
\eqn  
which is a quadratic  function of $h_{\mu\nu}$. Then, substituting Eqs.(\ref{eq2.5gd}) and (\ref{eq2.5ge}) back to  Eq.(\ref{eq2.5cac}), we find that the high-frequency part takes the form,
\bqn
\lb{eq2.5gf}
 R^{(1)}_{\mu\nu} &-& \frac{1}{2}\left(\gamma_{\mu\nu}R^{(1)} + h_{\mu\nu}R^{(0)}\right) \nb\\
 &+&   \epsilon \left<G^{(2)}_{\mu\nu}\right>^{\text{high}} =   \kappa\left<{\cal{T}}_{\mu\nu}\right>^{\text{high}},
\eqn
where 
\bqn
\lb{eq2.5gg}
  \left<G^{(2)}_{\mu\nu}\right>^{\text{high}} &\equiv&  G^{(2)}_{\mu\nu} -  \left<{G_{\mu\nu}}^{(2)}\right>_{\ell},\nb\\
 \left<{\cal{T}}_{\mu\nu}\right>^{\text{high}}  &\equiv&  {\cal{T}}_{\mu\nu}  -  \left<{\cal{T}}_{\mu\nu}\right>_{\ell}. 
 \eqn

On the other hand, from Eqs.(\ref{eq2.13ca})-(\ref{eq2.13cc}) we find that
\bqn
\lb{eq2.5gh}
&& G^{(0)}_{\mu\nu} \equiv R^{(0)}_{\mu\nu} -  \frac{1}{2}\gamma_{\mu\nu}R^{(0)} \simeq {\cal{O}}\left(\hat\gamma/\epsilon_c\right),\nb\\
&&  \left<{G_{\mu\nu}}^{(2)}\right>_{\ell} \simeq {\cal{O}}\left(h^2/\epsilon^2\right),\quad 
 T^{(0)}_{\mu\nu}   \simeq {\cal{O}}\left(\hat\gamma/\epsilon_c\right). ~~~~~~~
\eqn
Note that, after introducing the cosmological perturbation scale $L_c$, the leading order of $G^{(0)}_{\mu\nu}$ becomes  $G^{(0)}_{\mu\nu} \simeq \epsilon_c \hat{R}_{\mu\nu}
\simeq  {\cal{O}}\left(\hat\gamma/\epsilon_c\right)$, instead of $L^{-2}$ \cite{Garoffolo19}. The same is true for $T^{(0)}_{\mu\nu}$, as it can be seen from Appendix A. 
Then, from Eq.(\ref{eq2.5gd}) we find that each term has the following order, 
\bqn
\lb{eq2.5gi}
 {\cal{O}}\left(\hat\gamma/\epsilon_c\right) + {\cal{O}}\left(h^2\right) =  {\cal{O}}\left(\hat\gamma/\epsilon_c\right)  + \epsilon {\cal{O}}\left(\left< {{\cal{T}}_{\mu\nu}}\right>_{\ell}\right).  
\eqn
Therefore, to have the backreaction of the GWs to the background be negligible, so that the background spacetime $\gamma_{\mu\nu}$ is uniquely determined by $T^{(0)}_{\mu\nu}$,
i.e., 
\bqn
\lb{eq2.5gj}
{R_{\mu\nu}}^{(0)} - \frac{1}{2}\gamma_{\mu\nu} R^{(0)} = \kappa {T_{\mu\nu}}^{(0)},
\eqn
we must assume that 
\bqn
\lb{eq2.5gk}
h^2 \ll \frac{\hat\gamma}{\epsilon_c}, \\
\lb{eq2.5gl}
 \epsilon \cdot \left|\left< {{\cal{T}}_{\mu\nu}}\right>_{\ell}\right| \ll  \frac{\hat\gamma}{\epsilon_c}. 
\eqn
In addition, from Eq.(\ref{eq2.5gf}) we find that 
\bq
\lb{eq2.5gm}
 \epsilon \left<G^{(2)}_{\mu\nu}\right>^{\text{high}} \simeq {\cal{O}}\left(h^2/\epsilon\right).
\eq
Therefore, in order for the quadratic terms from $G^{(2)}_{\mu\nu}$ not to affect the linear terms of the leading orders 
 $h/\epsilon^{2}$ and $h/\epsilon^{1}$ in Eq.(\ref{eq2.5gf}), we must assume that
\bq
\lb{eq2.5gn}
|h| \ll  1.
\eq

With the above conditions, we find that Eq.(\ref{eq2.5gf}) can be written as
\bqn
\lb{eq2.5e}
\square\chi_{\alpha\beta} &+& \gamma_{\alpha\beta}\nabla^{\gamma}\nabla^{\delta}\chi_{\gamma\delta} -\nabla_{\alpha}\nabla^{\delta}\chi_{\beta\delta}
    -\nabla_{\beta} \nabla^{\delta}\chi_{\alpha\delta}\nb\\
    &+&  2{R_{\alpha\gamma  \beta  \sigma}}^{(0)} \chi^{\gamma\sigma}\nb\\
   &=& { \kappa\left({\cal{F}}_{\alpha\beta} -2 \left<{\cal{T}}_{\alpha\beta}\right>^{\text{high}}\right),}
\eqn
  {where
\bqn
\lb{eq2.5f}
        {\cal{F}}_{\alpha\beta} &\equiv & \frac{1}{\kappa}\Big\{{R^{\sigma}_{\;\;\alpha}}^{(0)} \chi_{\beta\sigma} + {R^{\sigma}_{\;\; \beta}}^{(0)} \chi_{\alpha\sigma}   - \chi_{\alpha\beta}R^{(0)} \nb\\
    &&+\gamma_{\alpha\beta}\chi^{\gamma\delta}{R_{\gamma\delta}}^{(0)} \Big\}  \nb\\
     &=&   \chi_{\beta\delta}{T^{\delta}_{\;\;\alpha}}^{(0)}  + \chi_{\alpha\delta}{T^{\delta}_{\;\;\beta}}^{(0)} +      \gamma_{\alpha\beta}\chi^{\gamma\delta}{T_{\gamma\delta}}^{(0)} \nb\\
  &&  - \frac{1}{2}\gamma_{\alpha\beta}\chi T^{(0)}.
\eqn
}From the above derivations, we can see that the linearized Einstein field equations (\ref{eq2.5e}) are valid only to the two leading orders, $\epsilon^{-2}$ and $\epsilon^{-1}$.
For orders higher than them, these equations are not applicable. This is particularly true for the zeroth-order of $\epsilon$. 

In addition, since $\epsilon^{-1}_c \ll \epsilon^{-1}$, we
find that in Eq.(\ref{eq2.5e}) the terms
  {
\bq
\lb{eq2.5eaa}
{\cal{F}}_{\alpha\beta}, \;\; 2{R_{\alpha\gamma  \beta  \sigma}}^{(0)} \chi^{\gamma\sigma} \simeq {\cal{O}}(\hat\gamma h/\epsilon_c) \ll  {\cal{O}}(h/\epsilon), 
\eq
which can  be also neglected, in comparing with  terms that are  orders of $\epsilon^{-2}$ or $\epsilon^{-1}$. However, in order to compare our results with the ones obtained in
\cite{Isaacson68a,Isaacson68b,LLSY10},  we shall keep them, and drop the corresponding terms only at the end of our calculations.}

\subsection{The Inhomogeneous  Universe}
\lb{II.C}

In this subsection, we shall give a very brief introduction over the flat FRW universe with its linear scalar and tensor perturbations, described by the metric (\ref{eq1.5}).
In terms of  the conformal coordinates $x^{\mu} = \left(\eta, x^i\right), (i = 1, 2, 3)$, 
we have
\bq
\lb{eq2.13a}
\bar{\gamma}_{\mu\nu} =  a^2(\eta) \eta_{\mu\nu},\quad \bar{\gamma}^{\mu\nu} =  a^{-2}(\eta) \eta^{\mu\nu},
\eq
with $\eta_{\mu\nu} = \text{diag}\left(-1, +1, +1, +1\right)$. The coordinate $\eta$ is related to the cosmic time via the relation,
$\eta = \int{\frac{dt}{a(t)}}$.

Following the standard process, we decompose the linear perturbations $\hat\gamma_{\mu\nu}$ into scalar, vector and tensor modes,
\bq
\lb{eq2.17}
\hat\gamma_{\mu\nu}
=a^2(\eta)\left(
    \begin{array}{c|c}
      -2\phi & \partial_i B - S_i  \\ \hline
      sym & -2\psi \delta_{ij} + 2 \partial_{ij}E +2 \partial_{(i}F_{j)} + H_{ij}
    \end{array}
  \right),
\eq
where
\bq
\lb{eq2.17a}
 \partial^{i}S_i =  \partial^{i}F_i = 0, \quad  \partial^{i}H_{ij} = 0 = H^{i}_{i},
 \eq
with    $ \partial^{i} \equiv \delta^{ij}\partial_{j}$ and $H^{i}_{j} \equiv \delta^{ik}H_{ kj}$. However, the vector mode will decay quickly with the expansion of the universe, and can be 
safely neglected \cite{Malik01,DB09}. Then, using the gauge transformations, as shown explicitly in Appendix A, we can always set
\bq
\lb{eq2.17b}
B = E = 0,
\eq
in which the gauge is completely fixed. This is often referred to as the Newtonian gauge, under which the gauge-invariant quantities defined in Eq.(\ref{eq2.27}) become, 
\bqn
\lb{eq2.17c}
 \Phi = \phi, \quad \Psi = \psi,\; (B = E = 0),
 \eqn
that is, in the  Newtonian gauge, the potentials $\phi$ and $\psi$ are equal to the gauge-invariant ones, $\Phi$ and $\Psi$.
Therefore,  with this gauge and ignoring the vector part,   we have
\bqn
\lb{eq2.17d}
\hat\gamma_{\mu\nu}
&=&a^2(\eta)\left(
    \begin{array}{c|c}
      -2\phi & 0  \\ \hline
      0 & H_{ij} -2\psi \delta_{ij}
    \end{array}
  \right),\nb\\
  \hat\gamma^{\mu\nu}
&=&a^{-2}(\eta)\left(
    \begin{array}{c|c}
      -2\phi & 0  \\ \hline
      0 & H^{ij} -2\psi \delta^{ij}
    \end{array}
  \right).
\eqn

In the rest of this paper, we shall restrict ourselves to this gauge.

 \section{Linearized Field Equations for   GWs   in  Inhomogeneous  Universe}
\renewcommand{\theequation}{3.\arabic{equation}} \setcounter{equation}{0}

In this section, we shall consider the field equations for $\chi_{\mu\nu}$ given by Eq.(\ref{eq2.5e}) in the inhomogeneous  cosmological background of Eq.(\ref{eq1.5}) with the Newtonian gauge
(\ref{eq2.17b}), by neglecting the vector perturbations, for which $\hat\gamma_{\mu\nu}$ and $\hat\gamma^{\mu\nu}$ are given by Eq.(\ref{eq2.17d}).

\subsection{Gauge Fixings  for GWs}

Before writing down these linearized field equations explicitly, let us first consider the gauge freedom for $\chi_{\mu\nu}$. At the end of the last section, we had considered the gauge transformations for the
cosmological perturbations, and had already used the 
gauge freedom,
\bq
\lb{eq2.6a}
\tilde{x}^{\mu} = x^{\mu} + \epsilon_c \zeta^{\mu},
\eq
to set $B = E = 0$ [cf. Eq.(\ref{eq2.17b})], the so-called Newtonian gauge, as shown explicitly in Appendix A. These choices  completely fix the gauge freedom for the cosmological perturbations.

 In this subsection, we shall consider another kind of gauge transformations  for the GWs, given by  
\bq
\lb{2.6}
\check{x}^{\alpha} = x^{\alpha} + \epsilon \xi^{\alpha},
\eq
where \footnote{In writing down the leading order of $ \xi_{\alpha}$, we had set the slowly-changing part that is of order one to zero, as it is irrelevant to the high frequency GWs considered here.}
\bq
\lb{2.7}
     \xi_{\alpha}   \simeq {\cal{O}}\left(\epsilon h\right), \;\;\;   \xi_{\alpha; \beta} \simeq {\cal{O}}\left(h \right), \;\;\;   \xi_{\alpha; \beta; \gamma} \simeq {\cal{O}}\left(h/\epsilon\right).
\eq
Since $\epsilon_c \gg \epsilon$,  we can see that to the first order of $\epsilon_c$, the background metric $\gamma_{\mu\nu}$ does not change under the 
coordinate transformations (\ref{2.6}), that is,
\bq
\lb{2.7a}
     \check\gamma_{\mu\nu} = \gamma_{\mu\nu}  + {\cal{O}}\left(\epsilon_c^2\right),
\eq
a property that is required for the transformations (\ref{2.6})  to be  the  gauge transformations only for the GWs. On the other hand, under the coordinate transformations (\ref{2.6}), we  have
\bqn
\lb{2.8}
\check{g}_{\mu\nu} &\equiv&  \check\gamma_{\mu\nu} + \epsilon \check{h}_{\mu\nu}  + {\cal{O}}\left(\epsilon^2\right) \nb\\
&=&     \gamma_{\mu\nu} + \epsilon \left(h_{\mu\nu} -  \xi_{\mu;\nu} - \xi_{\nu;\mu} \right) +  {\cal{O}}\left(\epsilon^2\right),
\eqn
 that is,
 \bq
 \lb{eq2.9}
 \check{h}_{\mu\nu} = h_{\mu\nu} -  2\xi_{(\mu;\nu)}.
 \eq
 Hence, we find
 \bqn
 \lb{eq2.10}
&& {{\check{R}}_{\alpha\beta\gamma\delta}}^{\;\;\;\;\;\;\;\; (1)}   - {R_{\alpha\beta\gamma\delta}}^{(1)} =  - {\cal{L}}_{\xi} {R_{\alpha\beta\gamma\delta}}^{(0)}  =  {\cal{O}}\left(h\hat\gamma/\epsilon_c\right),\nb\\
&& {{\check{R}}_{\alpha\beta}}^{\;\;\;\;(1)}   - {R_{\alpha\beta}}^{(1)} =  - {\cal{L}}_{\xi} {R_{\alpha\beta}}^{(0)}  =  {\cal{O}}\left(h\hat\gamma/\epsilon_c\right),
 \eqn
as can be seen from Eqs.(\ref{eq2.13ca})-(\ref{eq2.13cc}), and (\ref{2.7}), where ${\cal{L}}_{\xi}$ denotes the Lie derivative. 
Therefore, Eq.(\ref{eq2.5e}) is gauge-invariant only up to ${\cal{O}}\left(h\hat\gamma/\epsilon_c\right)$.
However, since $\epsilon^{-1}_c \ll \epsilon^{-1}$, terms that are order of $\epsilon^{-2}$ and $\epsilon^{-1}$ are still gauge-invariant, while 
the ones of order of $\epsilon^{0}$ are not. 
 This is because in the scale $\lambda$ the spacetime appears locally flat, and the curvature is locally gauge-invariant. Thus, provided that the following conditions hold,
\bq
\lb{eq2.10a}
 |h|,\; |\hat\gamma| \ll 1, \quad  \epsilon \ll \epsilon_c \ll 1,
\eq
the GW produced by an astrophysical source can be considered as a high-frequency GW, and their low-frequency components are negligible, so that
 the local-flatness behavior carries over to the case in which the background is even curved.

On the other hand, from the field equations (\ref{eq2.5e}) we can see that they will be considerably simplified, if we choose the Lorenz gauge,
\bq
\lb{eq3.0b}
\nabla^{\nu}\check{\chi}_{\mu\nu} = 0,  
\eq
where
\bqn
\lb{eq3.0c}
\check{\chi}_{\mu\nu} &\equiv& \check{h}_{\mu\nu} - \frac{1}{2}\gamma_{\mu\nu} \check{h}\nb\\
&=& \chi_{\mu\nu} - 2\nabla_{(\mu}\xi_{\nu)} + \gamma_{\mu\nu} \nabla_{\lambda}\xi^{\lambda},
\eqn
as it can be seen from Eq.(\ref{eq2.9}), where $\xi_{\mu} \equiv \gamma_{\mu\nu}\xi^{\nu}$. Then, we find that  the Lorenz gauge (\ref{eq3.0b}) requires,
\bqn
\lb{eq3.0d}
\Box\xi_{\mu} + {R^{(0)}}^{\nu}_{\mu} \xi_{\nu} = \nabla^{\nu}{\chi}_{\mu\nu}.
\eqn
Note that ${R^{(0)}}^{\nu}_{\mu} \xi_{\nu} \simeq {\cal{O}}(h\hat\gamma\epsilon/\epsilon_c) \ll {\cal{O}}(h/\epsilon)$, so {to the order of $\epsilon^{-1}$
it can be neglected.}
 Clearly, for any given $\chi_{\mu\nu}$  (with some proper continuous conditions \cite{Bernstein50}, which are normally assumed always  to exist.),
the above equation in general  has non-trivial solutions \cite{Isaacson68a}. 

In addition,  Eq.(\ref{eq3.0d}) does not completely fix the gauge. In fact, the gauge residual, 
\bq
\lb{eq3.0e}
\check{\check{x}}^{\alpha} = \check{x}^{\alpha} + \epsilon \varsigma^{\alpha},
\eq
exists, 
 for which the Lorenz gauge (\ref{eq3.0b}) still holds, 
 \bq
 \lb{eq3.0ea}
 \nabla^{\nu}\check{\check{\chi}}_{\mu\nu} = 0, 
 \eq
 as long as $ \varsigma^{\alpha}$
 satisfies the conditions,
\bq
\lb{eq3.0f}
\Box \varsigma_{\mu} + {R^{(0)}}^{\nu}_{\mu} \varsigma_{\nu} = 0.
\eq
Again, in this equation the term $ {R^{(0)}}^{\nu}_{\mu} \varsigma_{\nu} \simeq {\cal{O}}(h\epsilon\hat\gamma/\epsilon_c)$ is negligible, in comparing with the one $\Box \varsigma_{\mu}  
\simeq {\cal{O}}(h/\epsilon)$. 

An interesting question is that: can we use this gauge residual further to set 
\bq
\lb{eq3.0fa}
\check{\check{\chi}}_{0\mu} = 0.
\eq
To answer this question, we first note that if this is the case, $\varsigma_{\mu}$ must satisfy the additional conditions,   
\bq
\lb{eq3.0g}
\nabla_0\varsigma_{\nu} + \nabla_{\nu}\varsigma_{0}   - \gamma_{0\nu}  \nabla_{\alpha}{\varsigma^{\alpha}} = \check{\chi}_{0\nu}.
\eq
Clearly, for any given $\gamma_{\mu\nu}$ and $\check\chi_{\mu\nu}$ (again with certain regular conditions \cite{Bernstein50}), in general the above equation   has solutions.
However, we must remember that $\varsigma_{\nu}$ also needs to satisfy Eq.(\ref{eq3.0f}). To see if these conditions are consistent or not, let us take the covariant derivative $\nabla^{\mu}$ in both sides of
Eq.(\ref{eq3.0g}), which results in
\bqn
\lb{eq3.0h}
 \nabla_{\nu}\nabla_0\varsigma^{\nu} &+& \Box\varsigma_{0}   - \nabla_0\nabla_{\nu}   \varsigma^{\nu} \nb\\   
&=& \Box\varsigma_{0} +  {R^{(0)}}_{0\alpha}\varsigma^{\alpha} =  0 = \nabla^{\nu}\check{\chi}_{0\nu}.
\eqn
 Therefore, we conclude that {\it it is consistent to impose the Lorenz and spatial gauges  simultaneously, even when the background is curved}  \cite{Isaacson68a}.
 
 Finally, we note that the traceless condition
 \bq
 \lb{eq3.0i}
 \chi = 0,
 \eq
 was also introduced in \cite{Isaacson68a}. In fact, provided that the Lorenz gauge $\nabla^{\nu}\chi_{\mu\nu} = 0$ holds, from the field equations (\ref{eq2.5e}) we find
 {
 \bqn
\lb{eq2.5ea}
\square\chi + 2 {R_{\alpha\beta}}^{(0)} \chi^{\alpha\beta}
   = \kappa \gamma^{\alpha\beta}\left({\cal{F}}_{\alpha\beta} -2 \left<{\cal{T}}_{\alpha\beta}\right>^{\text{high}}\right).  ~~~
\eqn
 Note that the two terms  ${\cal{F}}$ and $2{R_{\gamma  \sigma}}^{(0)} \chi^{\gamma\sigma}$  are order of $h\hat\gamma/\epsilon_c$,
  as shown above, and can be dropped in comparing with terms of the order $h/\epsilon$. } Therefore,
 far from the source (${\cal{T}}_{\alpha\beta} = 0$), if the Lorenz gauge holds, one can also consistently impose the traceless gauge. 
 Together with the Lorenz and spatial gauges, it leads to the well-known  traceless-transverse (TT) gauge, frequently used when the  background is Minkowski \cite{MM16,DInverno03,PW14}. 
 
 It should be noted that in curved backgrounds the above three different gauge conditions can be imposed simultaneously only for high frequency GWs, and are valid only up to the order of $\epsilon^{-1}$
  \cite{Isaacson68a}. In other situations, when imposing them, one must pay  great cautions, as these constraints in general represent much more degrees than the four degrees of
 the gauge freedom that the general covariance normally allows.

\subsection{Field Equations for GWs}

To write down explicitly the field equations (\ref{eq2.5e}) for $\chi_{\mu\nu}$,  { and to make our expressions as much applicable as possible, in Appendix A, we only impose the spatial gauge,
\bq
\lb{eq3.8ch}
\chi_{0\mu} = 0, \; (\mu = 0, 1, 2, 3),
\eq
and then calculate each term appearing in  Eq.(\ref{eq2.5e}), before putting} them together to finally obtain the explicit expressions for each component of the field equations. In particular, the non-vanishing components of  
  {${\cal{F}}_{\alpha\beta}$ and $2{R_{\gamma\alpha\sigma\beta}}^{(0)}\chi^{\gamma \sigma}$ are given by Eqs.(\ref{eqA.2}) and (\ref{eqA.5a}),} while the ones of  $\Box\chi_{\alpha\beta}$ are given by Eqs.(\ref{eqA.7}) and (\ref{eqA.8}). The term
$\gamma_{\alpha\beta}  \nabla^{\gamma}\nabla^{\delta}\chi_{\gamma\delta}$ is given by Eqs.(\ref{eqA.10}) and (\ref{eqA.11}), while the one $\nabla_{\alpha}\nabla^{\delta}\chi_{\beta\delta}$ is given by Eq.(\ref{eqA.13}). 
Setting 
\bqn
\lb{eq3.9}
{\cal{G}}_{\alpha\beta} &\equiv& \square\chi_{\alpha\beta} +  \gamma_{\alpha\beta}\nabla^{\gamma}\nabla^{\delta}\chi_{\gamma\delta} -\nabla_{\alpha}\nabla^{\delta}\chi_{\beta\delta}
    -\nabla_{\beta} \nabla^{\delta}\chi_{\alpha\delta} \nb\\
    && +  2{R_{\alpha \gamma\beta  \sigma}}^{(0)} \chi^{\gamma\sigma},
\eqn
we find that the field equations (\ref{eq2.5e}) take the form,
\bq
\lb{eq3.10}
{\cal{G}}_{\alpha\beta} = \kappa\left({\cal{F}}_{\alpha\beta} -2 \left<{\cal{T}}_{\alpha\beta}\right>^{\text{high}}\right),  
\eq
where the non-vanishing components of ${\cal{G}}_{\alpha\beta}$ are given by Eqs.(\ref{eqA.15a}) - (\ref{eqA.15c}).

 \section{Geometrical optics approximation}  
\renewcommand{\theequation}{4.\arabic{equation}} \setcounter{equation}{0}

To study the propagation of GWs   in our inhomogeneous universe, let us first note that, when far away from the source that produces the GWs, we have ${\cal{T}}_{\mu\nu} = 0$. Then, 
Eq.(\ref{eq3.10}) reduces to,   
  {
\bq
\lb{eq4.1}
{\cal{G}}_{\alpha\beta} = \kappa{\cal{F}}_{\alpha\beta}, \; \left({\cal{T}}_{\mu\nu} = 0\right). 
\eq
 }
 Following Isaacson \cite{Isaacson68a} and Laguna et al \cite{LLSY10}, we consider the geometrical optics approximation, for  which we have
 {
 \bq
 \lb{eq4.2}
 \chi_{\alpha\beta} = \text{Re}\left(A_{\alpha\beta} e^{i\varphi/\epsilon}\right) =   \text{Re}\left(e_{\alpha\beta}{\cal{A}}e^{i\varphi/\epsilon}\right),
 \eq
 where $e_{\alpha\beta}$ denotes the polarization tensor with 
 \bq
 \lb{eq4.2aa}
 e^{\alpha\beta} e_{\alpha\beta}^* = 1, 
 \eq
 and 
 ${\cal{A}}$ and $\varphi$ are real and } characterize, respectively, the amplitude and phase of the GWs with $e^{\alpha\beta} \equiv \gamma^{\alpha\mu} \gamma^{\beta\nu}e_{\mu\nu}$. Note that in writing the above expression we made the change,
 $\varphi_I \rightarrow \varphi/\epsilon$, by following  Laguna et al \cite{LLSY10}, where $\varphi_I$ is the quantity used by Isaacson \cite{Isaacson68a}. 
 With this in mind, we can see that both the amplitude ${\cal{A}}$ and the phase $\varphi$ are slowly changing functions \cite{Isaacson68a}, 
  \bq
 \lb{eq4.2a}
\partial_{\alpha}\varphi  \simeq  {\cal{O}}(1),  \quad {A^{\alpha\beta}}_{; \gamma}  \simeq  {\cal{O}}(1).   
 \eq
 With the gauge
 (\ref{eq3.8ch}), we must set 
 \bq
 \lb{eq4.2b}
 A_{0 \beta} = 0 =  e_{0 \beta}. 
 \eq

Moreover, as shown in the last section, in addition to the spatial gauge, we can consistently impose  the Lorenz and traceless gauges,
\bq
\lb{eq4.2ba}
\nabla^{\nu}\chi_{\mu\nu} = 0, \quad \chi = 0.
\eq
Then,    from  Eqs.(\ref{eq4.2}) and (\ref{eq4.2b}) we find that the Lorenz gauge yields, 
\bq
\lb{eq4.2bb}
\nabla^{\nu}A_{\mu\nu} + \frac{i}{\epsilon} k^{\nu}A_{\mu\nu} = 0,
\eq
 where $k_{\alpha}\equiv \nabla_{\alpha}\varphi$ and $k^{\alpha}\equiv \gamma^{\alpha\beta}k_{\beta}$. Considering Eq.(\ref{eq4.2a}) we find that, to the leading order ($\epsilon^{-1}$), we have
 \bq
\lb{eq4.2cc}
k^{\nu} A_{\mu\nu} = 0  \quad \Rightarrow \quad k^{\nu} e_{\mu\nu} = 0.
\eq
 Therefore,  {\it the propagation direction of the GW is orthogonal to its polarization plane spanned by the bivector $e_{\mu\nu}$}. Note that the first term in Eq.(\ref{eq4.2bb}) is of order $\epsilon^0$, and should
 be discarded. Otherwise, it will lead to inconsistent results, as mentioned above. Therefore, in the rest of this paper we shall ignore such terms without further notifications.  See  \cite{Isaacson68a,LLSY10,Garoffolo19} for more details.  
 
 In addition, the traceless condition requires
 \bq
\lb{eq4.2bba}
\gamma^{\alpha\beta} e_{\alpha\beta} =   0.
\eq
Plugging Eq.(\ref{eq4.2})  into Eq.(\ref{eq3.10}) and considering Eq.(\ref{eq4.2a})  and the Lorenz gauge (\ref{eq4.2ba}),
we find that  the field equations to the orders of $\epsilon^{-2}$ and $\epsilon^{-1}$ are given, respectively, by 
 \bqn
  \lb{eq4.5a}
 &&  \epsilon^{-2}:    \quad k^{\mu}k_{\mu}A_{\alpha \beta} = 0, \\
      \lb{eq4.5b}
  &&  \epsilon^{-1}:     \quad k^{\mu}\nabla_{\mu} e_{\alpha \beta}+   \left(k^{\mu}\nabla_{\mu}\ln{\cal{A}} +\frac{1}{2}\nabla_{\mu}k^{\mu}\right)e_{\alpha \beta} = 0.\nb\\
    \eqn

 Since  $A_{\mu\nu} \not= 0$,  from Eq.(\ref{eq4.5a}) we find
 \bq
 \lb{eq4.5d}
  k^{\lambda}k_{\lambda} = 0.
 \eq
 Then, for such a null vector $k^{\mu}$, we can always define a curve $x^{\mu} = x^{\mu}(\lambda)$ by setting 
  \bq
 \lb{eq4.5e}
  \frac{dx^{\mu}(\lambda)}{d\lambda} \equiv  k^{\mu},
 \eq
where $\lambda$ denotes the affine parameter along the curve.  It is clear that such a defined curve is a null geodesics, 
   \bq
 \lb{eq4.5f}
k^{\lambda}\nabla_{\mu}k_{\lambda}  = k^{\lambda}\nabla_{\lambda}k_{\mu} = 0,    
 \eq
 as now we have $\nabla_{\mu}k_{\lambda} = \nabla_{\mu}\nabla_{\lambda}\varphi =   \nabla_{\lambda}\nabla_{\mu}\varphi = \nabla_{\lambda}k_{\mu}$, that is, 
  {\it GWs are always propagating along null geodesics in our inhomogeneous universe, even  when both the cosmological scalar and tensor
perturbations are all present, as long as the geometrical optics approximation are valid}. 
 
On the other hand,   
   Multiplying $e^{\alpha\beta}$ in both sides of Eq.(\ref{eq4.5b}) and taking Eq.(\ref{eq4.2aa}) into account, we find that 
  \bqn
  \lb{eq4.5h} 
       \frac{d}{d\lambda} \ln{\cal{A}}  + \frac{1}{2} \nabla_{\mu}k^{\mu} =  0,
   \eqn
  where $d/d\lambda \equiv k^{\nu}\nabla_{\nu}$.  Introducing the current $J^{\mu} \equiv {\cal{A}}^2 k^{\mu}$ of the gravitons moving along the null geodesics,
    the above equation can be written in the form,
    \bqn
  \lb{eq4.5i} 
        \nabla_{\mu}J^{\mu} =  0.
   \eqn
Therefore,  {\it the current of the gravitons moving along the null geodesics defined by $k^{\mu}$ is  conserved, even when
the primordial GWs (or cosmological tensor perturbations) are present ($H_{ij} \not= 0$)}. 

  Inserting Eq.(\ref{eq4.5h}) into Eq.(\ref{eq4.5b}), we find that
\bqn
\lb{eq4.5k}
 k^{\mu}\nabla_{\mu} e_{\alpha \beta} = 0.
\eqn
Thus, {\it the polarization bivector $e_{\alpha\beta}$ is still parallel-transported along the null geodesics, even  when the primordial GWs are present}. 

   {It should be to note that  {\it Eqs.(\ref{eq4.2bb})-(\ref{eq4.5k})  hold not only for the inhomogeneous universe, but also for any curved background, 
   as long as the geometrical optics approximation are applicable to the high frequency GWs}. For more detail, see \cite{MTW73}.}
   
  To study them further, we expand $\chi_{\mu\nu}$ in terms of $\epsilon_c$ as,
 \bq
\lb{eq3.8a}
 \hat\chi_{\mu\nu} =
 \chi^{(0)}_{\mu\nu} + \epsilon_c \chi^{(1)}_{\mu\nu} + {\cal{O}}\left(\epsilon_c^2\right),
 \eq
and then consider them order by order.

 \subsection{ GWs Propagating in Homogeneous and isotropic Background}

 To the zeroth-order of $\epsilon_c$, we have $\gamma_{\mu\nu} \simeq \bar{\gamma}_{\mu\nu} = a^2\eta_{\mu\nu}$, and 
\bq
\lb{eq4.3}
 \chi_{\mu\nu} \simeq \chi^{(0)}_{\mu\nu} + {\cal{O}}\left(\epsilon_c\right),
 \eq
 where we had set 
\bq
\lb{eq4.6a}
 {\chi}^{(0)}_{\mu\nu} \equiv  A^{(0)}_{\mu\nu} e^{i\varphi^{(0)}/\epsilon} = e^{(0)}_{\mu\nu}{\cal{A}}^{(0)}e^{i\varphi^{(0)}/\epsilon}. 
 \eq

Then, from Eqs.(\ref{eq4.5i}) and (\ref{eq4.5k}) we immediately obtain, 
\bqn
\lb{eq4.9}
&& \bar{\nabla}_{\nu}\left( {{\cal{A}}^{(0)}}^2 k^{(0) \nu}\right)    = 0, \\
\lb{eq4.11}
 && \frac{d }{d\lambda} e^{(0)}_{ij} = 0,
\eqn
where $k^{(0)}_{\mu} \equiv  \bar\nabla_{\mu}{\varphi^{(0)}} =  \left({\varphi^{(0)}}_{,\eta}, {\varphi^{(0)}}_{,i}\right)$, and
$k^{(0)\mu} \equiv \bar\gamma^{\mu\nu} k^{(0)}_{\nu}$.

 \subsection{Gravitational iSW  Effects}
 
 The derivation of the iSW effect in cosmology  { is based crucially on} the fact that the electromagnetic radiation propagating along null geodesics in the inhomogeneous universe. 
 Laguna et al \cite{LLSY10} took the advantage of the fact that GWs are also propagating along null geodesics and derived the gravitational  iSW effect for GWs 
when only the cosmological scalar perturbations are present ($H_{ij} = 0$). 
 In this subsection, we shall generalize their studies further to the case where both the cosmological
  scalar and tensor perturbations are present. As shown by Eq.(\ref{eq4.5d}), even when both of them are 
 present, the GWs produced by astrophysical sources are  still propagating along the null geodesics. Therefore, such a generalization is straightforward.  
 
 In particular,  let us first introduce the conformal metric $\tilde \gamma_{\mu\nu}$ by  
\bqn
\lb{eq4.13aaa}
d\tilde s^2 & =& \tilde \gamma_{\mu\nu} dx^{\mu}dx^{\nu} \equiv a^{-2}\gamma_{\mu\nu} dx^{\mu}dx^{\nu} =  - \left(1+ 2\epsilon_c \phi\right)d\eta^2 \nb\\
 &&  + \Big[\left(1- 2\epsilon_c \psi\right)\delta_{ij} + H_{ij}\Big]dx^idx^j.
 \eqn
 Since $\gamma_{\mu\nu}$ and $\tilde \gamma_{\mu\nu}$ are related to each other by a conformal transformation, so the null geodesics $x^{\mu}(\lambda)$ in the $\gamma_{\mu\nu}$ spacetime 
 is the same as $\tilde{x}^{\mu}(\tilde\lambda)$ in the  $\tilde \gamma_{\mu\nu}$ spacetime, where 
 \bq
 \lb{eq4.13bbb}
 d\lambda = a^2 d\tilde\lambda, \quad k^{\mu} = \frac{1}{a^{2}}\tilde{k}^{\mu},
 \eq
 and   $\tilde\lambda$ is the affine parameter of the null geodesics $\tilde{x}^{\mu}$ in the spacetime of    $\tilde \gamma_{\mu\nu}$.

\begin{figure}[htb]
\centering
\includegraphics[width=10cm]{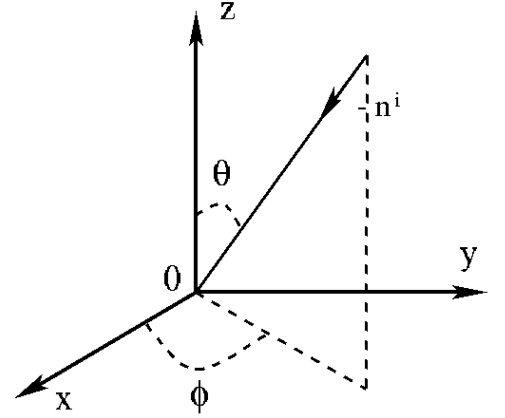}
\caption{A gravitational wave is propagating along the spatial direction $\tilde{k}^{(0) i} \equiv - n^{i}$ to the observer located at the origin.} 
\label{fig1}
\end{figure}

 The advantage of working with the metric $\tilde \gamma_{\mu\nu}$ is that the zeroth-order spacetime now becomes the Minkowski spacetime, and 
 the corresponding null geodesics are the straight lines, given by
 \bq
 \lb{eq4.14aaa}
 \frac{d\tilde{x}^{(0) \mu}(\tilde\lambda)}{d\tilde\lambda} \equiv \tilde{k}^{(0) \mu}.
 \eq 
 Thus, to simplify our  calculations, we shall work with  $\tilde \gamma_{\mu\nu}$.
In particular, to the zeroth-order of $\epsilon_c$, we have 
\bq
\lb{eq4.15aaa}
\tilde{k}^{(0) \mu}  = \left(1,- n^i\right), 
\eq
where  $\tilde k^{(0) i} \equiv -n^i$ represents the  spatial    direction of the GWs from the source propagating to the observer [cf. Fig. \ref{fig1}]. Then, from Eq.(\ref{eq4.9}) we find, 
\bq
\lb{eq4.16aaa}
\frac{d}{d\tilde\lambda}\ln\left(a {\cal{A}}^{(0)}\right) =   - \frac{1}{2}{\tilde{k}^{(0) \nu}}_{~~~~,\nu} = 0,
\eq
which implies that the quantity defined by
\bq
\lb{eq4.17aaa}
\mathcal{Q} \equiv {\cal{R}} {\cal{A}}^{(0)},
\eq
is  constant along the GW path, and will be determined by the local wave-zone source solution, where ${\cal{R}}  \equiv a r$ denotes the physical distance  between the observer and the source, 
while   $r$ denotes the comoving distance, given by
$r \equiv \sqrt{\left(x_e - x_r\right)^2 +\left(y_e - y_r\right)^2 + \left(z_e - z_r\right)^2}$, where $x_e^i \equiv (x_e, y_e, z_e)$ and $x_r^i \equiv (x_r, y_r, z_r)$ are the spatial locations of the source and observer, respectively.

In the following, we shall
set up the coordinates as follows \cite{LLSY10}: The observer is located at the origin with its proper time denoted by $\tau$ and world line $x^{\mu}(\tau)$. 
Denoting the time to receive the GW by $\tau_r$, this event will be
recorded as $x^{\mu}(\tau) = \left(\tau_{r}, \vec{o}\right)$. The emission time of the GW by an astrophysical source corresponds to the proper time $\tau_e$ of the observer with 
 $x^{\mu}(\tau) = \left(\tau_{e}, \vec{o}\right)$. Then, the GW will move along the null geodesics, described by
 $\tilde{x}^{\mu}(\tilde\lambda) = \tilde{x}^{(0) \mu}(\tilde\lambda) + \epsilon_c \tilde{x}^{(1) \mu}(\tilde\lambda)$, which corresponds to the wave vector 
 $\tilde{k}^{\mu}(\tilde\lambda)  = \tilde{k}^{(0) \mu}(\tilde\lambda) + \epsilon_c  \tilde{k}^{(1) \mu}(\tilde\lambda)$, where
 $ \tilde{x}^{(0) \mu}(\tilde\lambda) = \left(\tilde{\lambda}, (\tilde{\lambda}_r - \tilde{\lambda})n^i\right)$,
 and $\tilde{\lambda}_r$ is the moment when the GW arrives at the 
 origin with $\tau(\tilde{\lambda}_r) = \tau_{r}$.

 The effects of the scalar and tensor perturbations are manifested from the perturbations of the null geodesics. 
 Considering the fact $\tilde{\Gamma}^{(0) \mu}_{\nu\lambda} = 0$  in the $\tilde \gamma_{\mu\nu}$ spacetime,
 we find that, to the first-order of $\epsilon_c$, $\tilde{k}^{(1) \mu}(\tilde\lambda)$ is given by
 \bqn 
 \lb{eq4.16}
 \frac{d\tilde{k}^{(1) \mu}}{d\tilde\lambda} + \tilde{\Gamma}^{(1)\mu}_{\alpha\beta}  \tilde{k}^{(0) \alpha}  \tilde{k}^{(0) \beta} = 0,
 \eqn
 where $ \tilde{\Gamma}^{(1)\mu}_{\alpha\beta}$ denotes the Christoffel symbols of the first-order of $\epsilon_c$. 
 As mentioned previously, for the scalar perturbations, we shall not assume that $\psi = \phi$, that is, the trace of the anisotropic stress of the universe does not necessarily vanish, 
as shown by Eq.(\ref{eq2.29b}) in Appendix A. 
Then,  for $\mu = 0$ we find that
\bqn
\lb{eq4.20}    
    \frac{d}{d\lambda}\tilde{k}^{(1)0}&=&\partial_{\tau}(\phi+\psi)-2\frac{d\phi}{d\lambda}  -\frac{1}{2}n^{k}n^{l}\partial_{\tau}H_{kl}, ~~~~~~~~~
\eqn
where
\bq
\lb{eq4.20a}
\frac{d \Phi}{d\lambda} \equiv \left(\partial_{\tau} - n^i \partial_{i}\right)\Phi.
\eq
Thus,  integrating Eq.(\ref{eq4.20})  we find, 
\bqn
\lb{eq4.21}
   \tilde{k}^{(1)0}=&&-\left.\left(\phi+\psi\right)\right|^{\lambda}_{\lambda_e}+\frac{1}{2}n^kn^l H_{kl}|^{\lambda}_{\lambda_e}\nb \\
    &&-2\left.\phi\right|^{\lambda}_{\lambda_e}+I^{(s)}_{iSW}-\frac{1}{2}I^{(t)}_{iSW}, ~~~~~
\eqn
where $I^{(s)}_{iSW}$ represents the gravitational iSW effect due to the cosmological scalar perturbations, and was first calculated in \cite{LLSY10}. The new term $I^{(t)}_{iSW}$ is the gravitational integrated 
 effect due to the cosmological tensor perturbations. They are given, respectively, by
\bqn
\lb{eq4.22}
     I^{(s)}_{iSW}&\equiv& \int^{\lambda}_{\lambda_e}\partial_{\tau}(\phi+\psi)d\lambda',\\
 \lb{eq4.23}
    I^{(t)}_{iSW}&\equiv& n^kn^l\int^{\lambda}_{\lambda_e}\partial_{\tau}H_{kl}d\lambda'. 
\eqn
 
 On the other hand, the spatial components of the wave-vector are given by, 
\bqn
\lb{eq4.24}    
    \frac{d}{d\lambda}\tilde{k}^{(1)i}_{\parallel}&=& -n^i \Bigg[\partial_{\tau}(\phi+\psi)+\frac{d}{d\lambda}(\phi-\psi) \nb\\
    &&-\frac{1}{2}n^kn^l\Bigg(\frac{dH_{kl}}{d\lambda}+\partial_{\tau}H_{kl}\Bigg)\Bigg], ~~~~~~\\
\lb{eq4.25}
    \frac{d}{d\lambda}\tilde{k}^{(1)i}_{\perp}&=&-\perp^{ij}\bigg[\partial_{j}(\phi+\psi)-n^k\frac{dH_{jk}}{d\lambda}\nb\\
    &&-\frac{1}{2}n^kn^l\partial_jH_{kl}\bigg],
\eqn
where we had set $\tilde{k}^{(1)i}=\tilde{k}^{(1)i}_{\parallel}+\tilde{k}^{(1)i}_{\perp}$, with  the parallel component of the spatial wave-vector being defined by $\tilde{k}^{(1)i}_{\parallel}=n^in_j\tilde{k}^{(1)j}$,
 and the perpendicular component by $\tilde{k}^{(1)i}_{\perp}=\perp^i_j\tilde{k}^{(1)j}$. The projection operator $\perp^{i}_{j}$ is defined by $\perp^{i}_{j} = \delta^i_j - n^i n_j$, with
 $n_i \equiv \delta_{ik}n^k$. After integrations, the above two equations yield, 
\bqn
\lb{eq4.26}
    \tilde{k}^{(1)i}_{\parallel}&=&-n^i \Bigg[(\psi-\phi)\vert^{\lambda}_{\lambda_e}-\frac{1}{2}n^kn^lH_{kl}\vert^{\lambda}_{\lambda_e}\nb\\
    &&+I^{(s)}_{iSW}-\frac{1}{2}I^{(t)}_{iSW}\Bigg], \\
    \tilde{k}^{(1)i}_{\perp} &=&-\perp^{ij}\Bigg[\int^{\lambda}_{\lambda_e}\partial_j(\phi+\psi)d\lambda'-n^kH_{jk}\vert^{\lambda}_{\lambda_e}\nb\\
    &&-\frac{1}{2}n^kn^l\int^{\lambda}_{\lambda_e}\partial_jH_{kl}d\lambda'\Bigg].
\eqn
The GW phase is then given by, 
\bqn\lb{eq4.28}
    \frac{d\varphi}{d\lambda}=\phi+\psi-\frac{1}{2}n^kn^l\int^{\lambda}_{\lambda_e}H_{kl}d\lambda',
\eqn
which leads to
\bqn\lb{eq4.29}
    \delta\varphi=\varphi-\varphi_e=&&\int^{\lambda}_{\lambda_e}(\phi+\psi)d\lambda'\nb\\
    &&-\frac{1}{2}n^kn^l\int^{\lambda}_{\lambda_e}H_{kl}d\lambda'.
\eqn
The frequency of the GW is defined as $\omega=-u^{\mu}k_{\mu}$, where $u^{\mu}$ is the 4-velocity of the fluid of the universe, {given by Eqs.(\ref{eq2.20}) - (\ref{eq2.22}),}
 from which we find that  the ratio of receiving and emitting frequencies  is given by
\bqn
\lb{eq4.30}
    \frac{\omega_r}{\omega_e}=\frac{1-\Upsilon}{1+z},
\eqn
where $1+z\equiv a_r/a_e$, and
\bqn\lb{eq4.31}
    \Upsilon&\equiv&\phi\vert^{\lambda_r}_{\lambda_e}+v^i n_{i}+\frac{1}{2}n^kn^lH_{kl}\vert^{\lambda_r}_{\lambda_e}\nb\\
    &&-I^{(s)}_{iSW}\vert_{\lambda_r}+\frac{1}{2}I^{(t)}_{iSW}\vert_{\lambda_r}.
\eqn

In addition,  setting  {${\cal{A}} =  {\cal{A}}^{(0)}(1+\epsilon_c\xi)$, from Eq.(\ref{eq4.5h}) we find }
\bqn\lb{eq4.33}
    -2\frac{d\xi}{d\lambda}=\partial_{\tau}\tilde{k}^{(1)0}+\partial_i\tilde{k}^{(1)i}_{\parallel}+\partial_i\tilde{k}^{(1)i}_{\perp}+\tilde{\Gamma}^{(1)\mu}_{\mu\nu}\tilde{k}^{(0)\nu}, ~~~~
\eqn
where
\bqn
    \partial_{\tau}\tilde{k}^{(1)0}&=&\partial_{\tau}\left(-2\phi+I^{(s)}_{iSW}-\frac{1}{2}I^{(t)}_{iSW}\right),\nb\\
    \partial_{i}\tilde{k}^{(1)i}_{\parallel}&=&\frac{d}{d\lambda}\left(\psi-\phi+I^{(s)}_{iSW}\right)\nb\\
    && -\partial_{\tau}\left(\psi-\phi+I^{(s)}_{iSW}\right)\nb\\
    &&-\frac{1}{2}\frac{d}{d\lambda}\left(n^kn^lH_{kl}+I^{(t)}_{iSW}\right)\nb\\
    &&+\frac{1}{2}\partial_{\tau}\left(n^kn^lH_{kl}+I^{(t)}_{iSW}\right), \nb\\
    \partial_{i}\tilde{k}^{(1)i}_{\perp}&=&-\perp^{ij}\bigg[\int^{\lambda}_{\lambda_e}\partial_i\partial_j(\phi+\psi)d\lambda'\nb\\
    &&- n^k \partial_iH_{kj} -\frac{1}{2} n^kn^l\int^{\lambda}_{\lambda_e}\partial_i\partial_jH_{kl} d\lambda'\bigg],\nb\\
    \tilde{\Gamma}^{(1)\mu}_{\mu\nu}\tilde{k}^{(0)\nu}&=&\frac{d}{d\lambda}(\phi-3\psi).
\eqn
Notice that in the last term, there are no contributions from the tensor perturbations.  Collecting  all of this together, Eq.(\ref{eq4.33}) yields, 
\bqn\lb{eq4.35}
    -2\frac{d\xi}{d\lambda}&=&-\partial_{\tau}(\phi+\psi)+\frac{d}{d\lambda}\left(-2\psi+I^{(s)}_{iSW}\right)\nb\\
    &&-\perp^{ij}\int^{\lambda}_{\lambda_e}\partial_i\partial_j(\phi+\psi)d\lambda'\nb\\
    && +\frac{1}{2}n^kn^l\partial_{\tau}H_{kl} -\frac{1}{2}\frac{d}{d\lambda}\left(n^kn^lH_{kl}+I^{(t)}_{iSW}\right)\nb\\
    && +\perp^{ij}n^k\partial_iH_{jk}\nb\\
    &&+\frac{1}{2}\perp^{ij}n^kn^l\int^{\lambda}_{\lambda_e}\partial_i\partial_jH_{kl}d\lambda',
\eqn
which has the general solution,  
\bqn
\lb{eq4.36}
    \xi&=&-\psi\vert^{\lambda}_{\lambda_e}+\frac{1}{2}\perp^{ij}\int^{\lambda}_{\lambda_e}\int^{\lambda'}_{\lambda_e}\partial_i\partial_j(\phi+\psi)d\lambda'd\lambda''\nb\\
    &&-\frac{1}{2}n^k\Bigg[-\frac{1}{2}n^lH_{kl}\vert^{\lambda}_{\lambda_e}+\perp^{ij}\int^{\lambda}_{\lambda_e}\partial_iH_{jk}d\lambda'\nb\\
    &&+\frac{1}{2}\perp^{ij}n^l\int^{\lambda}_{\lambda_e}\int^{\lambda'}_{\lambda_e}\partial_i\partial_jH_{kl}d\lambda'd\lambda''\Bigg]. 
\eqn
 In terms of the gravitational tensorial  iSW effect defined by Eq.(\ref{eq4.23}), the above expression can be written in the form,
\bqn
\lb{eq4.37}
    \xi &=&\left. \left(\psi-\frac{1}{4}n^kn^lH_{kl}\right)\right|^{\lambda}_{\lambda_e} +\frac{1}{2}I^{(t)}_{iSW} \nb\\
    &&-\frac{1}{4}\perp^{ij}\int^{\lambda}_{\lambda_e}\int^{\lambda'}_{\lambda_e}\partial_i\partial_j\Bigg[n^kn^lH_{kl}  -2\left(\phi+\psi\right)\Bigg]d\lambda''d\lambda'\nb\\
    &&     -\frac{1}{2}n^k\int^{\lambda}_{\lambda_e}\partial^lH_{kl}d\lambda'.
\eqn

Combining all of our results together,  we are at the point to  construct the gravitational waveform through Eq.(\ref{eq4.2}), from which we find that   
\bqn
\lb{eq4.38}
h_{\mu\nu} &=& \chi_{\mu\nu} - \frac{1}{2}\chi \gamma_{\mu\nu} =  e_{\mu\nu}\tilde{h},\nb\\
    \tilde{h} &\equiv& \mathcal{A}e^{i\varphi} = \frac{(1+z)\mathcal{Q}}{d_L}(1+\xi)e^{i(\varphi_e+\delta\varphi)},
\eqn
where $\delta\varphi$ and $\xi$ are given, respectively, by Eqs.(\ref{eq4.29})  and (\ref{eq4.37}),   
and $d_L \equiv (1+z){\cal{R}}$ is the 
luminosity distance. Note that in writing the expression for the response function $\tilde h$ we had set $\epsilon = 1$.

For a binary system, we have \cite{PW14,LLSY10},    
\bqn
\lb{eq4.39}
\mathcal{Q} &=&\mathcal{M}_e\left(\pi f_e\mathcal{M}_e\right)^{2/3},\nb\\
\varphi_e &=& \varphi_c - \left(\pi f_e \mathcal{M}_e\right)^{-5/3},
\eqn
where $\mathcal{M}_e$ and $f_e$ denote, respectively,  the intrinsic chirp mass and frequency of the binary, and $\phi_c$ is the value of the phase at the merge, at which we have $f = \infty$. 
Therefore, the  function $\tilde h$ for a binary system  can be cast in the form,
\bqn
\lb{eq4.40}
    \tilde{h} =\frac{\mathcal{M}_r}{D_L}(\pi f_r \mathcal{M}_r)^{2/3}e^{i(\varphi_e+\delta\varphi)}, 
\eqn
where  the modified luminosity distance $D_L$ and the chirp mass ${\cal{M}}_r$ measured by the observer are given, respectively, by
\bqn
\lb{eq4.41}
D_L &\equiv& \frac{d_L}{1-\Upsilon-\xi}, \quad
\mathcal{M}_r \equiv \left(\frac{1+z}{1  -\Upsilon}\right) \mathcal{M}_e, 
\eqn
where $\Upsilon$ is given by Eq.(\ref{eq4.31}).

 \section{Conclusions}  
\renewcommand{\theequation}{5.\arabic{equation}} \setcounter{equation}{0}

In this paper, we have systematically studied GWs, which are first produced by some remote compact astrophysical sources, and then propagate in our inhomogeneous universe 
through cosmic distances before arriving at the detectors. Such GWs will carry valuable information  of both their sources and  the cosmological expansion and 
inhomogeneities of the universe, whereby a completely new window to explore our universe by using GWs is opened.  As the third generation (3G) detectors, such as
 the space-based ones,  LISA \cite{PAS17}, TianQin \cite{TianQin}, Taiji \cite{Taiji}, DECIGO \cite{DECIGO},   and the ground-based ones,  ET 
  \cite{ET} and  CE \cite{CE}, are able to detect GWs emitted from  such sources as far as at the redshift $z \simeq 100$ \cite{HE19}  {(See also Footnote 1),}  it is 
  very important and timely to carry out such studies systematically. Such studies were already initiated some years ago  \cite{LLSY10,BRBM17,BCST17} in the framework 
  of Einstein's theory, and more recently in scalar-tensor theories \cite{Garoffolo19,DFL20a,DFL20b,EZ20,EZ20}. 
 
 In this paper, in order to characterize effectively such systems,  we  first introduced three scales,
 $\lambda, \; L_c$ and $L$, which represent, respectively, the typical wavelength of the GWs, the scale of the cosmological perturbations, and the size of our 
 observable universe. For  GWs to be detected by the current and foreseeable (both ground- and space-based) detectors, in Sec. II we showed that the relation 
 \bq
 \lb{eq5.1}
 \lambda \ll L_c \ll L,
 \eq
  is always true, that is, such GWs can be well approximated as {\it high frequency GWs}, for which the general formulas were already developed by Isaacson more than half century ago 
  \cite{Isaacson68a,Isaacson68b}. 
  
  However,  Isaacson considered only the case where the background is vacuum, while in   \cite{LLSY10,BRBM17,BCST17} only the cosmological scalar perturbations were considered. 
  In this paper, we  considered the most general case in which  the background also includes the cosmological tensor perturbations. The inclusion of the latter is important, as now one of the main goals of 
  cosmological observations is the primordial GWs (the tensor perturbations) \cite{CMB-S4}. In the non-vacuum case, (in Sec. II) we showed explicitly that the conditions
  \bq
 \lb{eq5.2}
 \left|h_{\mu\nu}\right| \ll 1, \quad \epsilon \ll \epsilon_c  \ll 1,
 \eq
must hold,  in order for the backreaction of the GWs to the background to be neglected, and the linearized Einstein field equations given by
 Eq.(\ref{eq2.5e}) to hold, where the total metric of the spacetime  is expanded as
$g_{\mu\nu} = \gamma_{\mu\nu} + \epsilon h_{\mu\nu}$, with $\gamma_{\mu\nu} (\equiv \bar\gamma_{\mu\nu} + \epsilon_c \hat\gamma_{\mu\nu})$  representing the background.

In Sec. III, we considered the gauge choices, and found that the three different gauge conditions, {\it spatial, traceless, and Lorenz}, given  respectively by Eqs.(\ref{eq1.1a}) - (\ref{eq1.1c}),  can be still
imposed {\it simultaneously}, even when both the cosmological scalar and tensor perturbations are present, as long as the GWs can be approximated as the high-frequency GWs.
 However, by imposing only   the spatial gauge (\ref{eq1.1a}),  the linearized Einstein field equations  
 (\ref{eq2.5e})   are explicitly given  in Appendix B.  If  $\chi_{\mu\nu}$ is decomposed into two
   parts, 
   \bq
   \lb{eq5.2a}
   \chi_{\mu\nu} = \chi_{\mu\nu}^{(0)} + \epsilon_c \chi_{\mu\nu}^{(1)} + {\cal{O}}\left(\epsilon_c^2\right),
   \eq
    the field equations for $\chi_{\mu\nu}^{(1)}$ are given explicitly in Appendix C. 
   
   As an application of our general formulas, developed in Secs. II and III, in Sec. IV we studied the GWs by using the geometrical optics approximation,
   \bq
   \lb{eq5.3}
   \chi_{\alpha\beta} = e_{\alpha\beta}{\cal{A}}e^{i\varphi/\epsilon},
   \eq
   where   $e_{\alpha\beta}$ represents the polarization tensor,  
   ${\cal{A}}$ 
   and $\varphi$ denote, respectively,  the amplitude and phase of the GWs.
   We showed explicitly that even when both the cosmological scalar and tensor perturbations are present, {\it such GWs are still propagating along null geodesics, and the current of gravitons 
   moving along the null geodesics is conserved, and  the  polarization tensor is parallel-transported}, i.e., 
    \bq
   \lb{eq5.4}
   k^{\lambda}\nabla_{\lambda}k^{\mu} = 0, \quad k^{\lambda}\nabla_{\lambda}e_{\alpha\beta} = 0, \quad \nabla^{\lambda} J_{\lambda} = 0,
   \eq
   where $k_{\mu} \equiv \nabla_{\mu}\varphi, \; J_{\mu} \equiv {\cal{A}}^2 k_{\mu}$. In fact, {\it these are true for any curved background, provided that: (a) the GWs can be considered as high-frequency GWs; 
   and (b) the geometrical optics approximation are valid} \cite{MTW73}. 
   
   With these remarkable features, we calculated the effects of the cosmological scalar and tensor perturbations 
   on the amplitude ${\cal{A}}$ and phase $\varphi$, given by Eqs.(\ref{eq4.29}), (\ref{eq4.37}) and (\ref{eq4.38}). Restricting to GWs produced by a binary system,
   the effects of the cosmological perturbations, both scalar and tensor, on the luminosity distance and the chirp mass are given explicitly by Eq.(\ref{eq4.41}), which represent a natural generalization 
   of the results obtained in \cite{LLSY10,BRBM17,BCST17} to the case in which the cosmological tensor perturbations are also present. 
   
   {It should be noted that in cosmology the effects of the scalar and tensor perturbations of the homogeneous universe on the luminosity distances were studied in \cite{SW67,PC96} and 
   \cite{DD12}. Since in the geometrical optics approximations both GWs and electromagnetic waves (EWs) are all moving alone the null geodesics, the effects of  the cosmological scalar perturbations  
   on the luminosity distance of GWs carried out in  \cite{LLSY10} should be  the same as that obtained in  \cite{SW67,PC96} for EWs, while the ones of the cosmological tensor perturbations  
     carried out in this paper should be the same as that obtained in \cite{DD12} for EWs. However, the calculations of the GW phase are new. This is mainly due to the fact that the detection of GWs depends not only their amplitudes but also their phases \cite{MM16}, while the phases of EWs in cosmology do not play a significant role \cite{SD03}.}

   The applications  of our general formulas developed in this paper to other studies are immediate, including
   the gravitational analogue of the electromagnetic Faraday rotations \cite{Wang91,Wang20,HFZ19,FSW20}, and their detections by the space- and ground-based detectors.
   We wish to return to these important issues  in other occasions soon. 
   
   It would be also very important to extend such studies to  include the relations between the GWs and their sources,  high-order corrections to the geometrical optics  approximations, 
   and more interesting the non-high frequency GWs.

\begin{acknowledgments}

We would like very much to thank David Wand and Wen Zhao for valuable discussions.   This work was partially supported by 
the National Key Research and Development Program of China under  the Grant No. 2020YFC2201503, 
the National Natural Science Foundation of China under the grant Nos. 11675143, 11675145,  11705053,  11975203  and 12035005,
the Zhejiang Provincial Natural Science Foundation of China under Grant Nos. LR21A050001, LY20A050002, 
and the Fundamental Research Funds for the Provincial Universities of Zhejiang in China under Grant No. RF-A2019015.
The work of S.M.  was supported in part by Japan Society for the Promotion of Science Grants-in-Aid for Scientific Research No. 17H02890, 
 No. 17H06359, and by World Premier International Research Center Initiative, MEXT, Japan.
 J.F. and B.-W.L. acknowledge  the support from Baylor University through the Baylor University Physics graduate program.

\end{acknowledgments}

\section*{Appendix A: Decompositions of cosmological perturbations and gauge choice}
\renewcommand{\theequation}{A.\arabic{equation}} \setcounter{equation}{0}

Following   \cite{Malik01,DB09},   the linear perturbations $\hat\gamma_{\mu\nu}$ can be decomposed into scalar, vector and tensor modes,
and given explicitly by Eq.(\ref{eq2.17}).

The energy-momentum tensor ${T_{\mu\nu}}^{(0)}$ of a fluid  takes the form \cite{Malik01},
\bq
\lb{2.19}
{T^{\mu}_{\nu}}^{(0)} =  \left(\rho + p\right)u^{\mu}u_{\nu} + p \delta^{\mu}_{\nu} + \pi^{\mu}_{\nu},
\eq
where ${u}^{\mu}$ is the 4-velocity of the fluid,  $\rho$ and ${p}$ are its energy density and isotropic pressure, respectively, and $\pi^{\mu}_{\nu}$ is the anisotropic stress tensor, which
has only spatial components, i.e.,  $\pi^{\mu}_{0} = 0$.
Setting
\bqn
\lb{eq2.20}
\rho&=& \bar\rho + \epsilon_c \delta\rho, \quad p =  \bar{p} + \epsilon_c \delta p,\nb\\
u^{\mu} &=& \bar{u}^{\mu} +  \epsilon_c \delta{u}^{\mu},
\eqn
where $\bar{u}^{\mu} = a^{-1}\delta_{\eta}^{\mu}$ is the 4-velocity of the fluid of the homogenous and isotropic universe, and $\bar\rho$ and $\bar{p}$
are its energy density and isotropic pressure, respectively, we find that $ \delta{u}^{\mu}$ can be decomposed as
\bq
\lb{eq2.21}
\delta u^{\mu} = \frac{1}{a}\left(- \phi,  \partial^i v + v^i\right),
\eq
where $\partial_i v^i = 0$. Then, from  $u_{\mu} \equiv \gamma_{\mu\nu}u^{\nu} = \bar{u}_{\mu} +  \epsilon_c \delta{u}_{\mu}$, we find that
\bq
\lb{eq2.22}
\delta u_{\mu} = {a}\left(- \phi,  \partial_i v +  \partial_iB +  v_i - S_i\right),
\eq
which leads to $u^{\mu}u_{\mu} = -1 + {\cal{O}}\left(\epsilon_c^2\right)$, as expected.

On the other hand, setting $\pi^{j}_{i} = \epsilon_c \hat \pi^{j}_{i}$, similar to $\hat{\gamma}_{\mu\nu}$, the  anisotropic stress tensor $\hat \pi^{j}_{i}$ can be decomposed into
scalar, vector and tensor modes,
\bqn
\lb{eq2.23}
\hat \pi^{j}_{i} &=&   \left(\partial^j\partial_i - \frac{1}{3}\delta^j_{i}  \partial^2\right)\Pi + \frac{1}{2}\left(\partial_i\Pi^j+ \partial^j \Pi_i\right)
 + \Pi^j_{\;\; i},\nb\\
\eqn
where $\partial_i\Pi^i = 0 = \Pi^i_{\;\; i}$, $ \partial_j\Pi^j_{\;\; i}=0$, $\Pi^i \equiv \delta^{ik}\Pi_{k}, \; \Pi^i_{\;\; j} \equiv \delta^{ik}\Pi_{kj}$,  $\partial^2 \equiv \partial^i\partial_i$, etc.     Then, we find that
\bqn
\lb{eq2.24}
{T^0_{\;\;\;  0}}^{(0)}  &=& - \bar\rho - \epsilon_c \delta\rho, \nb\\
{T^0_{\;\;\; i}}^{(0)}  &=& \epsilon_c\left(\bar\rho + \bar{p}\right)\left[\partial_i (v + B) + v_i - S_i\right], \nb\\
{T^i_{\;\;\; 0}}^{(0)}  &=& - \epsilon_c\left(\bar\rho + \bar{p}\right)\left(\partial^i v  + v^i\right), \nb\\
{T^i_{\;\;\; j}}^{(0)} &=& \bar{p}\delta^i_j + \epsilon_c\left(\delta{p}\delta^i_j + \hat{\pi}^i_j\right).
  \eqn

  \subsection{ Gauge Transformations of Cosmological Perturbations}

 Considering the gauge transformations,
 \bq
\lb{2.25}
\tilde\eta =\eta +  \epsilon_c \zeta^0,\quad \tilde{x}^i = x^i + \epsilon_c\left(\partial^i\zeta + {\zeta}^i\right),
\eq
where $\partial_i {\zeta}^i = 0$, we find that
\bqn
\lb{2.26a}
\tilde\phi &=& \phi - {\cal{H}}\zeta^0 - {\zeta^0}', \;\;
\tilde\psi = \psi + {\cal{H}}\zeta^0, \nb\\
\tilde{B} &=& B + \zeta^0 - \zeta', \;\;
\tilde{E} = E- \zeta, \nb\\
\tilde{\delta\rho} &=& \delta\rho - \zeta^0 \bar{\rho}', \;\;
\tilde{\delta{p}} = \delta{p} - \zeta^0 \bar{p}', \nb\\
\tilde{v} &=& v + \zeta',\\
\lb{2.26b}
\tilde{F}_i &=& F_i - \zeta_i, \;\;
\tilde{S}_i = S_i + \zeta_i',\nb\\
\tilde{v}^i &=& v^i + {\zeta^i}', \\
\lb{2.26c}
\tilde{H}_{ij} &=& H_{ij},\quad
\tilde{\pi}^{i}_{j}   = \pi^{i}_{j},
\eqn
where ${\cal{H}} \equiv a'/a$ with $a' \equiv da/d\eta$. From the above gauge transformations we can see that the following quantities are gauge-invariant,
\bqn
\lb{eq2.27}
\Phi &\equiv& \phi + {\cal{H}}\left(B- E'\right) + \left(B- E'\right)',\nb\\
\Psi &\equiv& \psi - {\cal{H}}\left(B- E'\right),\nb\\
\Phi_i &\equiv& S_i + F_i'.
\eqn

On the other hand,  if we choose
$\zeta = E, \; \zeta^0 = E' - B$ and $\zeta_i = F_i$, we have
\bq
\lb{eq2.28}
\tilde{B} = \tilde{E} = 0, \quad \tilde{F}_i = 0,
\eq
in which the gauge is completely fixed. This is often referred to as the Newtonian gauge. Then, we are left with six scalars, $(\phi, \psi, v, \delta\rho, \delta{p}, \Pi)$, two vectors, $(S_i, v_i)$,  and two
tensors, $(H_{ij}, \Pi_{ij})$. However, the vector part decreases rapidly  with the expansion of the universe, so we can safely set them to zero \cite{Malik01,DB09},
\bq
\lb{eq2.29}
S_i  = F_i  = v_i = \Pi_i = 0.
\eq
 
 Then, for the scalar perturbations, there are six-independent equations, given, respectively, by  \cite{Malik01}, 
 \bqn
 \lb{eq2.29a}
&& \psi'' + 2 {\cal{H}}\psi' +  {\cal{H}}\phi' + \left(2 {\cal{H}}' +  {\cal{H}}^2\right)\phi \nb\\
&& ~~~~~~~~~~~~~~~~ = 4\pi G a^2\left(\delta p + \frac{2}{3}\nabla^2\Pi\right),\\
  \lb{eq2.29b}
&& \psi - \phi = 8\pi G a^2\Pi,\\
  \lb{eq2.29c}
&& 3 {\cal{H}}\left(\psi' +  {\cal{H}}\phi\right) - \nabla^2\psi = - 4\pi G a^2 \delta\rho,\\
  \lb{eq2.29d}
&& \psi' +  {\cal{H}}\phi = - 4\pi G a^2 \left(\bar\rho + \bar{p}\right) v, \\
  \lb{eq2.29e}
&& \delta\rho' + 3{\cal{H}}\left(\delta\rho + \delta p\right) = \left(\bar\rho + \bar{p}\right) \left(3\psi' - \nabla^2v\right),  ~~~~~\\
  \lb{eq2.29f}
&& \left[\left(\bar\rho + \bar{p}\right) v\right]' + \delta{p} + \frac{2}{3}\nabla^2\Pi \nb\\
&&  ~~~~~~~~~~~~~~~~ = - \left(\bar\rho + \bar{p}\right) \left(\phi + 4 {\cal{H}} v\right).
 \eqn
 Note that Eqs.(\ref{eq2.29a}) and (\ref{eq2.29b}) are obtained from the linearized (i, j)-components of the Einstein field equations, and
  Eqs.(\ref{eq2.29c}) and (\ref{eq2.29d}) are the energy and momentum constraints, while  Eqs.(\ref{eq2.29e}) and (\ref{eq2.29f}) are 
  obtained from the conservation of the energy-momentum tensor. 
  
For the tensor perturbations, we have 
\bqn
\lb{eq2.29g}
H_{ij}'' + 2 {\cal{H}} H_{ij}' - \nabla^2H_{ij} = 16\pi G a^2 \Pi_{ij},
\eqn
which is obtained from the equations $\delta {G^{(0)}}^{i}_{j} = \kappa\delta {T^{(0)}}^{i}_{j}$. 

It must be noted that in writing the linearized field equations, (\ref{eq2.29a}) - (\ref{eq2.29g}), we had implicitly assumed that the quadratic terms
$\epsilon_c^2 \hat R^{(2)}_{\mu\nu} (\hat\gamma) \simeq {\cal{O}}(\hat\gamma^2) \ll 1$, which is equivalent to
\bq
\lb{eq2.29h}
 \hat\gamma \ll 1,
\eq
where $\hat R^{(2)}_{\mu\nu} (\hat\gamma)$ is given by Eq.(\ref{eq2.4ab}) with the replacement $\left(h_{\mu\nu}, \nabla_{\alpha}\right) \rightarrow 
\left(\hat\gamma_{\mu\nu}, \bar\nabla_{\alpha}\right)$. 
Otherwise, these  quadratic terms cannot be neglected from the Einstein field equations for the background spacetimes,
\bq
\lb{eq2.29i}
\bar{G}_{\mu\nu}(\bar\gamma)  + \epsilon_c \hat{G}_{\mu\nu}(\hat\gamma) + \epsilon_c^2 \hat{G}^{(2)}_{\mu\nu}(\hat\gamma)  = \kappa T^{(0)}_{\mu\nu},
\eq
where 
\bqn
\lb{eq2.29j}
  \bar{G}_{\mu\nu}(\bar\gamma) &\simeq&  {\cal{O}}(1), \quad \epsilon_c \hat G_{\mu\nu} (\hat\gamma) \simeq {\cal{O}}(\hat\gamma/\epsilon_c), \nb\\
  \epsilon_c^2 \hat G^{(2)}_{\mu\nu} (\hat\gamma) &\simeq& {\cal{O}}(\hat\gamma^2),
\eqn
 as can be seen from Eq.(\ref{eq2.13cb}).

\section*{Appendix B: Field Equations for $\chi_{ij}$}
\renewcommand{\theequation}{B.\arabic{equation}} \setcounter{equation}{0}

In this Appendix, we shall calculate all the components of the quantities appearing in the field equations (\ref{eq3.10}) for $\chi_{\alpha\beta}$, by imposing only the
spatial gauge,
$$
\chi_{0\mu} = 0.
$$
In particular, to calculate  the non-vanishing components of the tensor ${\cal{G}}_{\alpha\beta}$, we first note that
 \begin{widetext}
\bqn
\lb{eqA.1}
&& \chi^{ij} \equiv \gamma^{i\mu} \gamma^{j\nu}\chi_{\mu\nu}  = \gamma^{ik} \gamma^{jl}\chi_{kl}
=  \frac{1}{a^2}\Big\{\delta^{ik}\delta^{jl} + \epsilon_c\left[4\psi \delta^{ik}\delta^{jl} - \left(\delta^{ik}H^{jl} + \delta^{jl}H^{ik}\right)\right]\Big\}\hat\chi_{kl},\nb\\
&& \gamma_{ij}\chi^{ij} = \hat\chi + \epsilon_c\left(2\psi\hat\chi - H^{kl}\hat\chi_{kl}\right),\quad  \chi \equiv \gamma^{\mu\nu}\chi_{\mu\nu} = \gamma^{ij}\chi_{ij} = \gamma_{ij}\chi^{ij},\nb\\
 && \gamma_{ij}\chi^{ik} \hat\pi^j_k = \Big[\hat\pi^{kl} + \epsilon_c\left(2\psi \hat\pi^{kl} - \hat\pi^k_mH^{ml}\right)\Big]\hat\chi_{kl}, \nb\\
  && \chi^{\gamma\sigma}{T_{\gamma\sigma}}^{(0)} - \frac{1}{2}\chi T^{(0)} = \frac{1}{2}\left(\bar\rho -\bar p\right) \hat\chi + \frac{1}{2}\epsilon_c\Big[\left(\bar\rho - \bar p\right)\left(2\psi \hat\chi -H^{kl}\hat\chi_{kl}\right)
  + \left(\delta\rho - \delta p\right) \hat\chi + 2 \hat\pi^{kl}\hat\chi_{kl}\Big],
\eqn
where $\hat\chi \equiv \delta^{ij}\hat\chi_{ij}$, $\chi_{ij} \equiv a^2 \hat\chi_{ij}$, $\hat\pi_{ij} \equiv \delta_{ik}\hat\pi^k_j$, etc.   Then, { from Eq.(\ref{eq2.5f}) we find that
\bqn
\lb{eqA.2}
{\cal{F}}_{00} &=& - \frac{a^2}{2}\Bigg\{\left(\bar\rho - \bar p\right)\hat\chi + \epsilon_c\Bigg[\left(\bar\rho - \bar p\right)\Big(2\left(\psi+\phi\right) \hat\chi - H^{kl}\hat\chi_{kl}\Big)
+ \left(\delta\rho - \delta p\right)\hat\chi + 2\hat\pi^{kl}\hat\chi_{kl}\Bigg]\Bigg\},  \nb\\
{\cal{F}}_{0i} &=& - a^2 \epsilon_c\left(\bar\rho + \bar p\right)\hat\chi_{ik}\partial^kv,  \nb\\
{\cal{F}}_{ij} &=& \frac{1}{2}a^2\Big[4\bar p \hat\chi_{ij} + \left(\bar\rho -\bar p\right) \hat\chi \delta_{ij}\Big]
+  \frac{1}{2}a^2 \epsilon_c\Bigg\{\Big[4\delta p \hat\chi_{ij} + \left(\delta\rho -\delta p\right) \hat\chi \delta_{ij}\Big]
+   \left(\bar\rho -\bar p\right)\left(\hat\chi H_{ij} - H^{kl}\hat\chi_{kl}\delta_{ij}\right) \nb\\
&&   + 2\left(\hat\pi^k_i\hat\chi_{jk} + \hat\pi^k_j\hat\chi_{ik} + \hat\pi^{kl}\hat\chi_{kl}\delta_{ij}\right)\Bigg\}.
\eqn
In addition,}   the non-vanishing (independent) components of the Riemann tensor,
 \bq
 \lb{A.3}
 {R_{\mu\nu\alpha\beta}}^{(0)} = \bar{R}_{\mu\nu\alpha\beta} + \epsilon_c \hat{R}_{\mu\nu\alpha\beta},
 \eq
 are given, respectively, by
\bqn
\lb{eqA.4}
 \bar{R}_{0i0j}&=& a^2\left({\cal{H}}^2 - \frac{a''}{a}\right)\delta_{ij}, \quad  \quad
\bar{R}_{minj}   =   a^2{\cal{H}}^2\left(\delta_{ij}\delta_{mn}- \delta_{in}\delta_{mj}\right),
\eqn
and
\bqn
\lb{eqA.5}
  \hat{R}_{0i0j}&=& a^2\Bigg\{\phi_{,ij}    + {\cal{H}}\phi' \delta_{ij}   +  \Bigg[\left(\psi'' + {\cal{H}}\psi'\right) + 2\left(\frac{a''}{a} - {\cal{H}}^2\right)\psi \Bigg] \delta_{ij}
                                 -\frac{1}{2}\Bigg[\left({H_{ij}}'' + {\cal{H}} {H_{ij}}'\right) +  2\left(\frac{a''}{a} - {\cal{H}}^2\right)H_{ij}\Bigg]\Bigg\},  \nb\\
   \hat{R}_{0ijk}&=&a^2\Big[{\cal{H}}\left(\phi_{,j}\delta_{ik} - \phi_{,k}\delta_{ij}\right) + \left(\psi_{,j}' \delta_{ik} - \psi_{,k}'\delta_{ij}\right)
                               + \frac{1}{2}\left(H_{ij,k}' - H_{ik,j}'\right)\Big],\nb\\
 \hat{R}_{ijkl}&=& -2a^2{\cal{H}}^2\phi\left(\delta_{ik}\delta_{jl} - \delta_{il}\delta_{jk}\right) - a^2\Big[\left(\delta_{jk} \psi_{,il} + \delta_{il} \psi_{,jk} - \delta_{ik} \psi_{,jl} - \delta_{jl} \psi_{,ik}\right)
                          + 2{\cal{H}}\left(\psi' + 2{\cal{H}}\psi\right)\left(\delta_{ik}\delta_{jl} - \delta_{il}\delta_{jk}\right)\Big]\nb\\
                     && + \frac{1}{2}a^2\Bigg\{\left(H_{jk,il} + H_{il,jk} - H_{ik,jl} - H_{jl,ik}\right) - {\cal{H}}\Big[\delta_{il}\left(H_{jk}' + 2{\cal{H}}H_{jk}\right) + \delta_{jk}\left(H_{il}' + 2{\cal{H}}H_{il}\right)\nb\\
                     && ~~~~~~~~~~ - \delta_{jl}\left(H_{ik}' + 2{\cal{H}}H_{ik}\right) - \delta_{ik}\left(H_{jl}' + 2{\cal{H}}H_{jl}\right)\Big]\Bigg\}.
\eqn
Hence, we find that   
\bqn
\lb{eqA.5a}
2{R_{0i0j}}^{(0)}\chi^{ij}&=&2\bigg(\frac{a''}{a}-\mathcal{H}^2\bigg)\hat{\chi}
+\epsilon_c\bigg\{2\left(\partial^i\partial^j\phi\right)\hat{\chi}_{ij}+2\mathcal{H}\phi'\hat{\chi}+2\big(\psi''+\mathcal{H}\psi'\big)\hat{\chi}-4\bigg(\frac{a''}{a}-\mathcal{H}^2\bigg)\psi\hat{\chi} 
                                               \nb\\
                                     &&~~~~~~~~~~~~~~~~~~~~~~~~~~~~  -\big({H^{ij}}''+\mathcal{H}{H^{ij}}'\big)\hat{\chi}_{ij} +2\bigg(\frac{a''}{a}-\mathcal{H}^2\bigg)H^{ij}\hat{\chi}_{ij}\bigg\}, \nb\\
2{R_{0jik}}^{(0)}\chi^{jk}&=&2\epsilon_c\Bigg\{\mathcal{H}\left[\left(\partial_i\phi\right)\hat{\chi}-\left(\partial^k\phi\right)\hat{\chi}_{ik}\right] +\left(\partial_i\psi'\right)\hat{\chi}-\left(\partial^k\psi'\right)\hat{\chi}_{ik} 
                                                +\frac{1}{2}\left[\left(\partial^k {H^j_i}'\right)-\left(\partial_i {H^{jk}}'\right)\right]\hat{\chi}_{jk}\Bigg\},  \nb\\
    2{R_{ikjl}}^{(0)} \chi^{kl}& =& 2\mathcal{H}^2\big(\delta_{ij}\hat{\chi}-\hat{\chi}_{ij}\big)+\epsilon_c\bigg\{4\mathcal{H}^2\phi\left(\hat\chi_{ij} - \hat{\chi}\delta_{ij}\right) +4\mathcal{H}\psi'\left(\hat{\chi}_{ij}- \hat{\chi} \delta_{ij}\right)\nb\\
    && 
    +2\Big[\left(\partial_i\partial_j\psi\right)\hat{\chi} +\left(\partial^k\partial^l\psi\right)\hat{\chi}_{kl} \delta_{ij}
    -\left(\partial^k\partial_i\psi\right)\hat{\chi}_{jk}-\left(\partial^k\partial_j\psi\right)\hat{\chi}_{ik}\Big]\nb\\
    && 
    -2\mathcal{H}^2H^{kl}\hat{\chi}_{kl} \delta_{ij} +2\mathcal{H}^2H_{ij}\hat{\chi}   
    -2\mathcal{H}{H^k_{(i}}'\hat{\chi}_{j)k} + \mathcal{H}{H_{ij}}'\hat{\chi} +\mathcal{H}{H^{kl}}'\hat{\chi}_{kl}  \delta_{ij} \nb\\
    && +\left(2\partial_{(i}\partial^lH^k_{j)}-\partial^k\partial^lH_{ij}-\partial_i\partial_jH^{kl}\right)\hat{\chi}_{kl}\bigg\}. 
\eqn

 On the other hand,  similar to the above expression, writing $\square\chi_{\alpha\beta}$     in the form,
\bqn
\lb{eqA.6}
\square\chi_{\alpha\beta} &\equiv& \bar\square\chi_{\alpha\beta} + \epsilon_c\hat\square\chi_{\alpha\beta},
\eqn
we find they are given, respectively,  by
\bqn
\lb{eqA.7}
    \bar\square\chi_{00}&=&2 {\cal{H}}^2 \hat{\chi}, \quad
    \bar\square\chi_{0i} = -2{\cal{H}}\partial^j\hat{\chi}_{ij},   \quad
 \bar\square\chi_{ij} = -{\hat{\chi}}_{ij}''-2{\cal{H}}{\hat{\chi}}_{ij}' +\partial^2\hat{\chi}_{ij} +2 {{\cal{H}}^2}\hat{\chi}_{ij},
\eqn
    and
\bqn
\lb{eqA.8}
 \hat\square\chi_{00}&=&
 - 2 {\cal{H}}\Big[2\left(\psi' -  {\cal{H}}\psi\right)\hat\chi - \left({H^{ij}}' -  {\cal{H}}H^{ij}\right)\hat\chi_{ij}\Big],\nb\\
 \hat\square\chi_{0i}&=&
\left(\partial^j\phi'\right)\hat\chi_{ij} + 2\left(\partial^j\phi\right)\left(\hat\chi_{ij}' +  {\cal{H}}\hat\chi_{ij}\right) + \left(\partial^j\psi'\right)\hat\chi_{ij} + 2\left(\psi' - 2 {\cal{H}}\psi\right) \partial^j\hat\chi_{ij}
+ 2{\cal{H}}\left[\left(\partial^j\psi\right)\hat\chi_{ij} - \left(\partial_i\psi\right)\hat\chi\right]\nb\\
&& - \left({H^{jk}}'- 2{\cal{H}}H^{jk}\right)\partial_k\hat\chi_{ij} + {\cal{H}}\left(\partial_iH^{jk}\right)\hat\chi_{jk},\nb\\
  \hat{\square}\chi_{ij}&=&
2\phi \hat\chi_{ij}''   + \left(\phi' + 4 {\cal{H}}\phi\right)\hat\chi_{ij}' + \left(\partial^k\phi\right)\partial_k\hat\chi_{ij} -4 {\cal{H}}^2\phi \hat\chi_{ij}\nb\\
&& + 2\psi\partial^2\hat\chi_{ij} + 4\partial_{(i}\psi\partial^k\hat\chi_{j)k} + 3\left(\partial^k\psi\right)\partial_k\hat\chi_{ij} - 4\left(\partial^k\psi\right)\partial_{(i}\hat\chi_{j)k}
+ 2\partial^k\partial_{(i}\psi\hat\chi_{j)k} + 2\left(\partial^2\psi\right) \hat\chi_{ij} \nb\\
&& -2 \partial_{(i}\partial^k\psi\hat\chi_{j)k} - \psi'{\hat\chi_{ij}}' - 2\left(\psi'' + 4{\cal{H}}\psi'\right)\hat\chi_{ij}\nb\\
&& -H^{kl}\partial_k\partial_l\hat\chi_{ij} - 2\partial^lH^k\hsp_{(i}\partial_l\hat\chi_{j)k} -2 \partial_{(i}H^{kl}\partial_l\hat\chi_{j)k} + 2\partial^kH_{(i}\hsp^l\partial_l\hat\chi_{j)k}
 + 2H^k\hsp_{(i}\hsp'\hat\chi_{j)k}' +H^k\hsp_{(i}\hsp''\hat\chi_{j)k} \nb\\
&&
   +4\mathcal{H}H^k\hsp_{(i}\hsp'\hat\chi_{j)k} - \partial^2H^k\hsp_{(i}\hat\chi_{j)k},
  \eqn
where $2\partial^k\partial_{(i}\psi\hat\chi_{j)k} \equiv   \left(\partial^k\partial_{i}\psi\right)\hat\chi_{jk} +   \left(\partial^k\partial_{j}\psi\right)\hat\chi_{ik}$, that is, the partial derivative acts only to the first function.
The same is true for other terms, for example, $ 2\partial^lH^k\hsp_{(i}\partial_l\hat\chi_{j)k} \equiv  \left(\partial^lH^k\hsp_{i}\right)\partial_l\hat\chi_{jk} +  \left(\partial^lH^k\hsp_{j}\right)\partial_l\hat\chi_{ik}$.

On the other hand, defining
\bqn
\lb{eqA.9}
{\cal{G}}_{\alpha\beta}^{(1)} \equiv \gamma_{\alpha\beta}  \nabla^{\gamma}\nabla^{\delta}\chi_{\gamma\delta},
   \eqn
we find that
\bqn
\lb{eqA.10}
   {\cal{G}}_{00}^{(1)} &=&  - {\cal{G}}^{(1)}_0 - \epsilon_c\left(2\phi {\cal{G}}^{(1)}_0 + {\cal{G}}^{(1)}_1\right), \quad
    {\cal{G}}_{0i}^{(1)} =   {\cal{G}}_{i0}^1 =  0, \nb\\
     {\cal{G}}_{ij}^{(1)} &=& \delta_{ij} {\cal{G}}^{(1)}_0 + \epsilon_c\left[\delta_{ij}\left({\cal{G}}^{(1)}_1 - 2\psi{\cal{G}}^{(1)}_0\right) + H_{ij} {\cal{G}}^{(1)}_0\right],
  \eqn
where
 \bqn
 \lb{eqA.11}
 {\cal{G}}^{(1)}_0 &\equiv& {\cal{H}}\hat\chi' + \left(\frac{a''}{a} + {\cal{H}}^2\right)\hat\chi + \partial^i\partial^j\hat\chi_{ij}, \nb\\
  {\cal{G}}^{(1)}_1&\equiv&   -2{\cal{H}}\phi \hat\chi' - \left[2\left(\frac{a''}{a} + {\cal{H}}^2\right)\phi + {\cal{H}} \phi' \right]\hat\chi + 2\left(\partial^i\phi\right)\left(\partial^j\hat\chi_{ij}\right)
   + \left(\partial^i\partial^j\phi\right)\hat\chi_{ij}\nb\\
     &&-   \left(\psi' - 2{\cal{H}}\psi\right) \hat\chi' - \left[\psi'' + 3   {\cal{H}} \psi' - \partial^2\psi - 2\left(\frac{a''}{a} +  {\cal{H}}^2\right)\psi \right] \hat\chi
     + (\pp^i\psi)\partial_i\hat\chi + 4 \psi  \partial^i\partial^j\hat\chi_{ij} -  \left(\partial^i\partial^j\psi\right)\hat\chi_{ij}\nb\\
      &&+   \frac{1}{2}\Bigg\{\left({H^{ij}}'  - 2{\cal{H}}H^{ij}\right) \hat\chi_{ij}' +  \left[{H^{ij}}''   - 2\left(\frac{a''}{a} +  {\cal{H}}^2\right)H^{ij}
      - \partial^2 H^{ij}\right]\hat\chi_{ij}  - 4H^{ik}\partial_k\partial^j\hat\chi_{ij} -     \left(\partial_k H^{ij}\right)\left(\partial^k\hat\chi_{ij}\right) \nb\\
      && ~~~~~~~
      -  2 \left(\partial^i H^{jk}\right)\left(\partial_k\hat\chi_{ij}\right)  \Bigg\}.
  \eqn

On the other hand, defining
\bqn
\lb{eqA.12}
{\cal{G}}_{\alpha\beta}^{(2)} \equiv  \nabla_{\alpha}\nabla^{\delta}\chi_{\beta\delta},
  \eqn
we find that it has  the following non-vanishing components,
  \bqn
\lb{eqA.13}
{\cal{G}}_{00}^{(2)} &=&  -\frac{a'}{a}\hat\chi'-\left(\frac{a''}{a} - 2\mathcal{H}^2\right)\hat\chi  + \epsilon_c\Bigg[\mathcal{H} \phi'\hat\chi-\left(\pp^i\phi\right) \pp^k\hat\chi_{ik}
           +  \left(\psi' -2\mathcal{H}\psi\right)\hat\chi'   + \left(\psi'' -3\mathcal{H}\psi'\right)\hat\chi -2\left(\frac{a''}{a}-2\mathcal{H}^2\right)\psi\hat\chi \nb\\
          && -\frac{1}{2}\left({H^{ij}}' - 2{\cal{H}}H^{ij}\right)\hat\chi_{ij}'  -\frac{1}{2}\left({H^{ij}}'' - 3\mathcal{H}{H^{ij}}'\right)\hat\chi_{ij}
        +\left(\frac{a''}{a}-2\mathcal{H}^2\right)H^{ij}\hat\chi_{ij} \Bigg], \nb\\
{\cal{G}}_{0i}^{(2)} &=&   \pp^k\hat\chi_{ik}'-\mathcal{H}\pp^k\hat\chi_{ik} + \epsilon_c\Bigg[\left(\pp^j\phi\right) \hat\chi_{ij}' + \left(\pp^j\phi' -\mathcal{H}\pp^j\phi\right)\hat\chi_{ij}+ \mathcal{H}\left(\pp_i\phi\right)\hat\chi\nb\\
         && + 2\psi\pp^k\hat\chi_{ik}' -\left(\pp^k\psi\right)\hat\chi_{ik}' + \left(3\psi' - 2\mathcal{H} \psi\right) \pp^k\hat\chi_{ik}  + \left(\pp_i\psi\right) \hat\chi'
        - \left(\pp^k\psi' - \mathcal{H}\pp^k\psi\right)\hat\chi_{ik} +\left(\pp_i\psi'-\mathcal{H}\pp_i\psi\right) \hat\chi \nb\\
        &&-H^{jk}\pp_k\hat\chi_{ij}'   -\left({H^{jk}}' -  \mathcal{H}H^{jk}\right) \pp_k\hat\chi_{ij}  -\frac{1}{2}{H^j_i}'\pp^k\hat\chi_{jk}  -\frac{1}{2}H^{jk}\hsp_{,i}\hat\chi_{jk}'
        -\frac{1}{2}\left({H^{jk}}' - \mathcal{H}H^{jk}\right)_{,i}\hat\chi_{jk} \Bigg], \nb\\
 {\cal{G}}_{i0}^{(2)} &=&     -\mathcal{H}\left(\pp^k\hat\chi_{ik} + \pp_i\hat\chi\right) + \epsilon_c\Bigg[\mathcal{H}\left(\pp_i\phi\right) \hat\chi  -\mathcal{H}\left(\pp^k\phi\right)\hat\chi_{ik}  \nb\\
 && +\left(\psi' -2\mathcal{H}\psi\right)\left(\pp^k\hat\chi_{ik} +   \pp_i\hat\chi\right) +  \mathcal{H}\left(\pp^k\psi\right)\hat\chi_{ik} + \left(\psi' -3{\cal{H}}\psi\right)_{,i} \hat\chi  \nb\\
             && -\frac{1}{2}\left({H^{jk}}' - 2{\cal{H}}H^{jk}\right)\pp_i\hat\chi_{jk}  -\frac{1}{2}{H^j_i}' \pp^k\hat\chi_{jk}   +\mathcal{H}H^{jk}\pp_k\hat\chi_{ij}
               -\frac{1}{2} \left({H^{jk}}' - 3{\cal{H}}H^{jk}\right)_{,i}\hat\chi_{jk}\Bigg], \nb\\
 {\cal{G}}_{ij}^{(2)} &=& \pp_i\pp^k\hat\chi_{jk}+\mathcal{H}^2\hat\chi \delta_{ij}+ \epsilon_c\Bigg[\left(\pp^k\phi\right)\pp_i\hat\chi_{jk} + \left(\pp_i\pp^k\phi\right)\hat\chi_{jk} -2\mathcal{H}^2\phi\hat\chi\delta_{ij} \nb\\
&& -\pp_i\left[\left(\pp^k\psi\right)\hat\chi_{jk}\right] +\pp_i\left[\left(\pp_j\psi\right)\hat\chi\right]   +2\pp_i(\psi\pp^k\hat\chi_{jk})
 +\left(\pp_i\psi\right)\pp^k\hat\chi_{jk}+\left(\pp_j\psi\right)\pp^k\hat\chi_{ik}-\left(\pp^k\psi\right) \pp^l\hat\chi_{kl}\delta_{ij}  -2\mathcal{H}\psi'\hat\chi\delta_{ij}  \nb\\
&& -\pp_i(H^{kl}\pp_l\hat\chi_{jk})-\frac{1}{2}\pp_i\left[\left(\pp_jH^{kl}\right)\hat\chi_{kl}\right]   +\frac{1}{2}\left(\pp^kH_{ij}-H^k\hsp_{i,j}-H^k\hsp_{j,i}\right)\pp^l\hat\chi_{kl}  \nb\\
&&
+\frac{1}{2}\mathcal{H}\left({H^{kl}}' - 2{\cal{H}}H^{kl}\right)\hat\chi_{kl}\delta_{ij}  +\frac{1}{2} \mathcal{H}\left(H_{ij}' + 2\mathcal{H}H_{ij}\right) \hat\chi\Bigg].
  \eqn
 Note that  ${\cal{G}}_{\alpha\beta}^{(2)}$ is not symmetric, ${\cal{G}}_{\alpha\beta}^{(2)} \not= {\cal{G}}_{\beta\alpha}^{(2)}$, as can be seen from its definition given by Eq.(\ref{eqA.12}).
 
 Finally, defining ${\cal{G}}_{\alpha\beta}$ as
 \bqn
\lb{eqA.14}
{\cal{G}}_{\alpha\beta} &\equiv& \square\chi_{\alpha\beta} +  \gamma_{\alpha\beta}\nabla^{\gamma}\nabla^{\delta}\chi_{\gamma\delta} -\nabla_{\alpha}\nabla^{\delta}\chi_{\beta\delta}
    -\nabla_{\beta} \nabla^{\delta}\chi_{\alpha\delta}  +  2{R_{\alpha \gamma\beta  \sigma}}^{(0)} \chi^{\gamma\sigma},
\eqn
we find that its non-vanishing components are given by
\bqn
\lb{eqA.15a}
    {\cal{G}}_{00}&=& \mathcal{H}\hat{\chi}'-\bigg(\frac{a''}{a}+\mathcal{H}^2   \bigg)\hat{\chi}-\partial^i\partial^j\hat{\chi}_{ij} \nonumber \\
    &&+\epsilon_c\Bigg\{\mathcal{H}\phi'\hat{\chi}+\left(\partial^i\partial^j\phi\right)\hat{\chi}_{ij}-2\phi\left(\partial^i\partial^j\hat{\chi}_{ij}\right) 
    +\bigg[\psi''+7\mathcal{H}\psi'
    -2\bigg(\frac{a''}{a}+\mathcal{H}^2\bigg)\psi-\partial^2\psi\bigg]\hat{\chi}-\big(\psi'-2\mathcal{H}\psi\big)\hat{\chi}' \nb \\
    && +\left(\partial^i\partial^j\psi\right)\hat{\chi}_{ij}-4\psi\left(\partial^i\partial^j\hat{\chi}_{ij}\right)-\left(\partial^k\psi\right)\partial_k\hat{\chi}   -\frac{1}{2} \bigg[\left({H^{ij}}'' - \partial^2H^{ij}\right) 
    + 4 \mathcal{H}{H^{ij}}'-2\bigg(\frac{a''}{a}+\mathcal{H}^2\bigg)H^{ij}\bigg]\hat{\chi}_{ij} \nonumber \\
    && +\frac{1}{2}\big({H^{ij}}'-2\mathcal{H}H^{ij}\big)\hat{\chi}'_{ij}+2H^{i}_k(\partial^k\partial^j\hat{\chi}_{ij})+\frac{1}{2}(\partial^kH^{ij})\partial_k\hat{\chi}_{ij}+(\partial^iH^{jk})\partial^k\hat{\chi}_{ij}\Bigg\}, \\
    \lb{eqA.15b}
    {\cal{G}}_{0i}&=& \mathcal{H}\partial_i\hat{\chi}-\partial^j\hat{\chi}'_{ij} \nonumber \\
    &&+ \epsilon_c\Bigg\{\left(\partial^j\phi\right)\hat{\chi}'_{ij}+2\mathcal{H}\left(\partial^j\phi\right)\hat{\chi}_{ij} 
    -2\psi'(\partial^j\hat{\chi}_{ij})+2\mathcal{H}(\partial^i\psi)\hat{\chi}-2\psi(\partial^j\hat{\chi}'_{ij})+(\partial^j\psi)\hat{\chi}'_{ij}-(\partial_i\psi)\hat{\chi}'-(\psi'-2\mathcal{H}\psi)\partial_i\hat{\chi} \nb\\
    &&-\mathcal{H}(\partial_i{H}^{jk})\hat{\chi}_{jk}+\left({\partial^kH^j_i}'\right)\hat{\chi}_{jk}+{H^j_i}'\left(\partial^k\hat{\chi}_{jk}\right)+H^{jk}\left(\partial_k\hat{\chi}'_{ij}\right)  
    +\frac{1}{2}(\partial_iH^{jk})\hat{\chi}'_{jk}+\frac{1}{2}{H^{jk}}'(\partial_i\hat{\chi}_{jk})-\mathcal{H}H^{jk}(\partial_i\hat{\chi}_{jk})\Bigg\},\nb\\
    \\
    \lb{eqA.15c}
    {\cal{G}}_{ij} &=& -\hat{\chi}''_{ij}+\partial^2\hat{\chi}_{ij}-2\mathcal{H}\hat{\chi}'_{ij}+\mathcal{H}\delta_{ij}\hat{\chi}'+\delta_{ij}\bigg(\frac{a''}{a}+\mathcal{H}^2\bigg)\hat{\chi}+\partial^k\partial^l\hat{\chi}_{kl}\delta_{ij}-\partial_i\partial^k\hat{\chi}_{jk}-\partial_j\partial^k\hat{\chi}_{ik} \nonumber \\
    &&+\epsilon_c\bigg\{2\phi\hat{\chi}''_{ij}+\big(\phi'+4\mathcal{H}\phi\big)\hat{\chi}'_{ij}-\bigg[\mathcal{H}\phi'+2\bigg(\frac{a''}{a}+\mathcal{H}^2\bigg)\phi\bigg]\delta_{ij}\hat{\chi}-2\mathcal{H}\phi\delta_{ij}\hat{\chi}' \nonumber \\
    &&-(\partial_j\partial^k\phi)\hat{\chi}_{ik}-(\partial_i\partial^k\phi)\hat{\chi}_{jk}+(\partial^k\phi)\partial_k\hat{\chi}_{ij}+2(\partial^k\phi)\partial^l\hat{\chi}_{kl}\delta_{ij}-(\partial^k\phi)\partial_i\hat{\chi}_{kj}-(\partial^k\phi)\partial_j\hat{\chi}_{ik}+(\partial^k\partial^l\phi)\hat{\chi}_{kl}\delta_{ij} \nonumber \\
    &&+2\big(\partial^2\psi-\psi''-2\mathcal{H}\psi'\big)\hat{\chi}_{ij}+\big(\partial^{2}\psi-\psi''-3\mathcal{H}\psi'\big)\hat{\chi}\delta_{ij}-\psi'\big(\hat{\chi}'_{ij}+\hat{\chi}'\delta_{ij}\big)+2\psi\partial^2\hat{\chi}_{ij} \nonumber \\
    &&+\partial^k\psi\big(3\partial_k\hat{\chi}_{ij}+\partial_k\hat{\chi}\delta_{ij}-\partial_i\hat{\chi}_{jk}-\partial_j\hat{\chi}_{ik}+2\partial^l\hat{\chi}_{kl}\delta_{ij}\big)+2\psi\partial^k\partial^l\hat{\chi}_{kl}\delta_{ij}+(\partial^k\partial^l\psi)\hat{\chi}_{kl}\delta_{ij} \nonumber \\
    &&-4\partial_{(i}\psi\partial^k\hat{\chi}_{j)k}-2\partial_{(i}\partial^k\psi\hat{\chi}_{j)k}-4\psi\partial_{(i}\partial^k\hat{\chi}_{j)k}-2\partial_{(i}\psi\partial_{j)}\hat{\chi}\nonumber \\
    &&+\frac{1}{2}\big({H^{kl}}'-2\mathcal{H}H^{kl}\big)\hat{\chi}'_{kl}\delta_{ij}+\bigg[\frac{1}{2}{H^{kl}}''-\bigg(\frac{a''}{a}+\mathcal{H}^2\bigg)H^{kl}\bigg]\hat{\chi}_{kl}\delta_{ij}+{H^k_{(i}}''\hat{\chi}_{j)k}+2\mathcal{H}{H^k_{(i}}'\hat{\chi}_{j)k} \nonumber \\
    &&+2{H^k_{(i}}'\hat{\chi}'_{j)k}+\bigg[\mathcal{H}\hat{\chi}'+\bigg(\frac{a''}{a}+\mathcal{H}^2\bigg)\hat{\chi}+\partial^k\partial^l\hat{\chi}_{kl}\bigg]H_{ij}-2H^{jk}\partial_k\partial^m\hat{\chi}_{lm}\delta_{ij}-\frac{1}{2}\partial^2H^{kl}\hat{\chi}_{kl}\delta_{ij}-\frac{1}{2}\partial_mH^{kl}\partial^m\hat{\chi}_{kl}\delta_{ij} \nonumber \\
    &&-2\partial^lH^k_{(i}\partial_l\hat{\chi}_{j)k}+2\partial^kH^l_{(i}\partial_l\hat{\chi}_{j)k}-\partial^2H^k_{(i}\hat{\chi}_{j)k}+\partial_{(i}H^{kl}\partial_{j)}\hat{\chi}_{kl}+\partial_{(i}H^k_{j)}\partial^l\hat{\chi}_{kl}+2\partial_{(i}\partial^lH^k_{j)}\hat{\chi}_{kl}-\partial^k\partial^lH_{ij}\hat{\chi}_{kl} \nonumber \\
    &&-\partial^kH_{ij}\partial^l\hat{\chi}_{kl}+2H^{kl}\partial_{(i}\partial_l\hat{\chi}_{j)k}-\partial^kH^{ml}\partial_m\hat{\chi}_{kl}\delta_{ij}-H^{kl}\partial_k\partial_l\hat{\chi}_{ij}\bigg\}.
\eqn

 \end{widetext}

 \section*{Appendix C:  Field Equations to the First-order of $\epsilon_c$}
\renewcommand{\theequation}{C.\arabic{equation}} \setcounter{equation}{0}

Following Eq.(\ref{eq3.8a}),  
we write $\hat\chi_{\alpha\beta}$ in the form,
\bq
\lb{C.1}
\hat\chi_{\alpha\beta} \simeq \hat\chi^{(0)}_{\alpha\beta} + \epsilon_c  \hat\chi^{(1)}_{\alpha\beta} +  {\cal{O}}\left(\epsilon_c^2\right), 
\eq
where to the zeroth-order, the TT gauge 
\bq
\lb{C.2}
\hat\chi^{(0)}_{0 \beta} = 0, \quad \hat\chi^{(0)} = 0, \quad \partial^i \hat\chi^{(0)}_{ij} = 0,
\eq
will be chosen. But, to the first order, we shall not impose the traceless and Lorenz gauge conditions. The only gauge that now we  choose is 
\bq
\lb{C.3} 
\hat\chi_{0\beta}^{(1)} = 0.
\eq
 With this gauge choice, to the first-order of $\epsilon_c$, the non-vanishing components of the tensor ${\cal{G}}_{\alpha\beta}$ given by Eqs.(\ref{eqA.15a})-(\ref{eqA.15c}) yield, 
\begin{widetext}
\bqn
\lb{C.4a}
    \mathcal{G}^{(1)}_{00}&=&\mathcal{H}\hat{\chi}'^{(1)} -\bigg(\frac{a''}{a}+\mathcal{H}^2   \bigg)\hat{\chi}^{(1)}-\partial^i\partial^j\hat{\chi}_{ij}^{(1)}+\left[\partial^i\partial^j\left(\phi+\psi\right)\right]\hat{\chi}^{(1)}_{ij}   + \hat{\mathcal{G}}^{(1)}_{00},\\
\lb{C.4b}
    \mathcal{G}^{(1)}_{0i}&=&\mathcal{H}\partial_i\hat{\chi}^{(1)}-\partial^j\hat{\chi}'^{(1)}_{ij} + \hat{\mathcal{G}}^{(1)}_{0i},\\
 \lb{C.4c}
    \mathcal{G}^{(1)}_{ij}&=&
    -\hat{\chi}''^{(1)}_{ij}+\partial^2\hat{\chi}_{ij}^{(1)}-2\mathcal{H}\hat{\chi}'^{(1)}_{ij}+\mathcal{H}\delta_{ij}\hat{\chi}'^{(1)}+\delta_{ij}\bigg(\frac{a''}{a}+\mathcal{H}^2\bigg)\hat{\chi}^{(1)}\nb\\
    && +\partial^k\partial^l\hat{\chi}^{(1)}_{kl}\delta_{ij}-\partial_i\partial^k\hat{\chi}^{(1)}_{jk}-\partial_j\partial^k\hat{\chi}_{ik}^{(1)} + \hat{\mathcal{G}}^{(1)}_{ij},
\eqn
where
\bqn
\lb{C.4aa}
    \hat{\mathcal{G}}^{(1)}_{00}&=&    -\frac{1}{2} \bigg[\left({H^{ij}}'' - \partial^2H^{ij}\right)   + 4 \mathcal{H}{H^{ij}}'-2\bigg(\frac{a''}{a}+\mathcal{H}^2\bigg)H^{ij}\bigg]\hat{\chi}^{(0)}_{ij} \nonumber \\
    && +\frac{1}{2}\left({H^{ij}}'-2\mathcal{H}H^{ij}\right)\hat{\chi}'^{(0)}_{ij}+\frac{1}{2}\left(\partial^kH^{ij}\right)\partial_k\hat{\chi}^{(0)}_{ij}+\left(\partial^iH^{jk}\right)\partial^k\hat{\chi}^{(0)}_{ij}, \\  
    \lb{C.4bb}
    \hat{\mathcal{G}}^{(1)}_{0i} &=& +\left(\partial^j\phi\right)\hat{\chi}'^{(0)}_{ij}+2\mathcal{H}\left(\partial^j\phi\right)\hat{\chi}^{(0)}_{ij}+\left(\partial^j\psi\right)\hat{\chi}'^{(0)}_{ij}\nb\\
    &&-\mathcal{H}\left(\partial_iH^{jk}\right)\hat{\chi}^{(0)}_{jk}+\left(\partial^kH'^{j}_i\right)\hat{\chi}^{(0)}_{jk}+H^{jk}\left(\partial_k\hat{\chi}'^{(0)}_{ij}\right)+\frac{1}{2}H'^{jk}\left(\partial_i\hat{\chi}^{(0)}_{jk}\right)
      -\mathcal{H}H^{jk}\left(\partial_i\hat{\chi}^{(0)}_{jk}\right), \\
    \lb{C.4cc}
    \hat{\mathcal{G}}^{(1)}_{ij} &=& +2\phi\hat{\chi}''^{(0)}_{ij}+\left(\phi'+4\mathcal{H}\phi\right)\hat{\chi}'^{(0)}_{ij}-\left(\partial_j\partial^k\phi\right)\hat{\chi}^{(0)}_{ik}-\left(\partial_i\partial^k\phi\right)\hat{\chi}^{(0)}_{jk}+\left(\partial^k\phi\right)\partial_k\hat{\chi}^{(0)}_{ij}
    -\left(\partial^k\phi\right)\partial_i\hat{\chi}^{(0)}_{kj}\nonumber\\
    &&-\left(\partial^k\phi\right)\partial_j\hat{\chi}^{(0)}_{ik}+\left(\partial^k\partial^l\phi\right)\hat{\chi}^{(0)}_{kl}\delta_{ij}\nonumber\\
    &&+2\left(\partial^2\psi-\psi''-2\mathcal{H}\psi'\right)\hat{\chi}^{(0)}_{ij}-\psi'\hat{\chi}'^{(0)}_{ij}+2\psi\partial^2\hat{\chi}^{(0)}_{ij}+\partial^k\psi\left(3\partial_k\hat{\chi}^{(0)}_{ij}-\partial_i\hat{\chi}^{(0)}_{jk}-\partial_j\hat{\chi}^{(0)}_{ik}\right)
    +\left(\partial^k\partial^l\psi\right)\hat{\chi}^{(0)}_{kl}\delta_{ij}\nonumber\\
    &&+\frac{1}{2}\left({H^{kl}}'-2\mathcal{H}H^{kl}\right)\hat{\chi}'^{(0)}_{kl}\delta_{ij}+\left[\frac{1}{2}{H^{kl}}''-\left(\frac{a''}{a}+\mathcal{H}^2\right)H^{kl}\right]\hat{\chi}^{(0)}_{kl}\delta_{ij}+{H^k_{(i}}''\hat{\chi}^{(0)}_{j)k}+2\mathcal{H}{H^k_{(i}}'\hat{\chi}^{(0)}_{j)k} \nonumber \\
    &&+2{H^k_{(i}}'\hat{\chi}'^{(0)}_{j)k}-\frac{1}{2}\partial^2H^{kl}\hat{\chi}^{(0)}_{kl}\delta_{ij}-\frac{1}{2}\partial_mH^{kl}\partial^m\hat{\chi}^{(0)}_{kl}\delta_{ij}-2\partial^lH^k_{(i}\partial_l\hat{\chi}^{(0)}_{j)k}+2\partial^kH^l_{(i}\partial_l\hat{\chi}^{(0)}_{j)k}-\partial^2H^k_{(i}\hat{\chi}^{(0)}_{j)k} \nonumber \\
    &&+\partial_{(i}H^{kl}\partial_{j)}\hat{\chi}^{(0)}_{kl}+2\partial_{(i}\partial^lH^k_{j)}\hat{\chi}^{(0)}_{kl}-\partial^k\partial^lH_{ij}\hat{\chi}^{(0)}_{kl}+2H^{kl}\partial_{(i}\partial^l\hat{\chi}^{(0)}_{j)k}-\partial^kH^{ml}\partial_m\hat{\chi}^{(0)}_{kl}\delta_{ij}. %
\eqn

 \end{widetext}

\end{document}